\documentclass{aa}  
\usepackage{graphicx}
\usepackage{txfonts}

\newcommand{\kms}{\mbox{km~s$^{-1}$}}
\newcommand{\vlsr}{\mbox{$\varv_{\rm lsr}$}}

\newcommand{\oho}{$^2\Pi_{1/2}$}
\newcommand{\oht}{$^2\Pi_{3/2}$}
\newcommand{\uchii}{UCHII}

\begin{document}

   \title{SOFIA/GREAT observations of OD and OH rotational lines towards high-mass star forming regions}

   \author{T. Csengeri 
          \inst{1}
          \and
          F. Wyrowski
           \inst{2}
          \and
         K. M. Menten
          \inst{2}
          \and
        H. Wiesemeyer
          \inst{2}
          \and R. G\"usten
          \inst{2}
          \and J. Stutzki
          \inst{3}
         \and S. Heyminck
          \inst{2}
           \and Y. Okada
          \inst{3}
                }

   \institute{Laboratoire d'astrophysique de Bordeaux, Univ. Bordeaux, CNRS, B18N, all\'ee Geoffroy Saint-Hilaire, 33615 Pessac, France
              \email{timea.csengeri@u-bordeaux.fr}
                \and
                Max-Planck-Institut f\"ur Radioastronomie,
              Auf dem H\"ugel 69, 53121 Bonn, Germany
              \and
              I. Physikalisches Institut der Universit\"at zu K\"oln, Z\"ulpicher Strasse 77, 50937 K\"oln, Germany
                  }

   \date{Received  ; accepted  }

  \abstract
  {Only recently, OD,  the deuterated isotopolog of hydroxyl, OH, has become accessible in the interstellar medium; spectral lines from both species have been observed in the supra-Terahertz and far infrared regime. 
  Studying variations of the OD/OH abundance amongst different types of sources can deliver key information on the formation of water, H$_2$O.} 
   {With observations of rotational lines of OD and OH towards  13 Galactic high-mass star forming regions, we aim to constrain the  {OD abundance and infer the} deuterium fractionation of OH  in their molecular envelopes. For the best studied source in our sample, G34.26+0.15, we were able to perform detailed radiative transfer modelling to investigate the OD abundance profile in its inner envelope. }
   {We used the Stratospheric Observatory for Infrared Astronomy (SOFIA) to observe the \oht\ $J=5/2-3/2$ ground-state transition of OD at 1.3\,THz ($215~\mu$m)  {{and the rotationally excited  OH line at 1.84\,THz ($163~\mu$m). We also used published high-spectral-resolution SOFIA data of the OH ground-state transition at} 2.51\,THz  ($119.3~\mu$m).} }
   {Absorption from the \oht\ OD $J=5/2-3/2$ ground-state transition is prevalent in the dense clumps  surrounding active sites of high-mass star formation. 
   Our modelling suggests that part of the absorption arises from the denser inner parts, while the bulk of it as seen with SOFIA originates in the outer, cold layers of the envelope for which our constraints on the molecular abundance  suggest a strong enhancement in deuterium fractionation.  
   We find a weak negative correlation between the OD abundance 
   and the bolometric luminosity to mass ratio,  
 an evolutionary indicator, suggesting a slow decrease of OD abundance with time. A comparison with HDO shows a similarly high deuterium fractionation for the two species in the cold envelopes, which is of the order of $0.48$\% {for the best studied source, G34.26+0.15}.}
   {Our results are consistent with chemical models that favour rapid exchange reactions to form OD in the dense cold gas. Constraints on the OD/OH ratio in the inner envelope could further elucidate the water and oxygen chemistry near young high-mass stars.}



   \maketitle
%

\section{Introduction}

The hydroxyl radical, OH, is a key molecule to our understanding of interstellar chemistry, in particular to constrain chemical pathways leading to the formation of water, 
a prime molecule of interest for star- and planet formation \citep[for a review see][]{Dishoeck2013}. Tracing its deuterated form, OD, promises  to place additional constraints on  formation routes through the establishment of the deuterium fractionation within protostellar envelopes.

The OH radical has been observed in various environments, first through its radio lines corresponding to transitions between rotational energy levels split by the magnetic hyperfine structure (hfs) of the molecule (\citealp[see e.g.][]{Weinreb1963} and \citealp{Weaver1964}). \citet{Cesaroni1991} discussed observations of multiple hfs transitions and modelled their excitation, which generally is \textit{\emph{not}} governed by local thermodynamic equilibrium (LTE). As shown in their energy level diagrams in Fig.\,\ref{fig:energy}, the rotational transitions of OH (and those of OD) lie in the supra-THz regime at $>$ 1.3\,THz and are only accessible from space and with airborne facilities. 
The first observations of rotational transitions of OH were performed with the Kuiper Airborne Observatory \citep{Storey1981, Melnick1987, Betz1989} and have been followed by the Infrared Space Observatory \citep{Ceccarelli1998,Giannini2001,Nisini2002}, all with modest spectral resolution of tens of \kms.
Only recently, observations of rotational transitions of OH have become accessible at higher spectral resolution with the Heterodyne Instrument for the Far-Infrared aboard Herschel (HIFI) \citep{Wampfler2011, Wampfler2013} and the German Receiver for Astronomy at Terahertz Frequencies (GREAT) on SOFIA \citep{Csengeri2012, Wiesemeyer2012, Parise2012, Wiesemeyer2016}.
With this receiver, the ground-state rotational line of its deuterated form, OD, was first  detected in absorption towards the envelope of the low-mass protostar  IRAS 16293$-$2422 \citep{Parise2012}. 

The OH molecule is a key component in the water chemistry pathway. At low temperatures, both species are intimately related, forming in the gas phase via dissociative recombination of the oxonium ion, $\mathrm{H_3O^+}$ {(\citealp{Neau2000,Jensen2000} and references therein). The latter is synthesised through a sequence of molecular hydrogen abstraction reactions \citep{Dalgarno1994} and is therefore sensitive to the molecular gas fraction.
At  high temperatures, OH forms through gas phase reactions from the reaction between atomic oxygen and hydrogen molecules, and through successive hydrogenation is converted into water \citep{Charnley1997,Harada2010,vanDishoeck2014}. The reaction of OH + H$_2$ {$\rightleftharpoons$} H$_2$O + H,  which is endothermic by~920 K, is sensitive to the ratio of atomic O and H and thus depends strongly on the local conditions, such as the temperature and UV field. OH is also a by-product of H$_2$O photodissociation, making it a  sensitive tracer of the radiation field. This, for example, results in the high OH abundances in the dense molecular envelopes of (ultra)compact HII regions \citep[see e.g.\,][]{Cesaroni1991, Csengeri2012}. The formation of OH is also dependent on the cosmic ray ionisation rate and the gas density \citep{Wakelam2010}.

Since OH is chemically linked to the formation of water, to measure the OD/OH ratio in star forming regions presents a promising tool to constrain the formation conditions of water. For the first time, we can  observationally constrain this ratio, which is expected to probe the chemical formation pathway for these molecules, as deuterium enhancement (i.e. fractionation) is strongly dependent on the chemical route and its formation conditions \citep[e.g.][]{Croswell1985, Roberts2002a}. Since deuterium is of solely primordial origin \citep{Peebles1966} and is destroyed in stars through fusion to $^3$He, a better understanding of its fractionation into OD may ultimately help to set constraints on star formation history.

Making use of the capabilities of the SOFIA/GREAT receiver, we  investigated the OD/OH ratio towards a sample of 13 Galactic high-mass, star forming regions.   
Most of our targets are well-known Galactic massive star-forming regions subject to extensive studies (e.g.\,\citealp{vanderTak2006,Wyrowski2016,Konig2017}).
They cover a range of evolutionary stages, characterized by clumps hosting embedded ultra-compact HII ({\uchii}) regions and internally heated hot molecular cores, as well as   quiescent massive sources. Our classification of the evolutionary stage of the sample is based on the type of embedded sources associated with them. In short, sources associated with compact radio emission from the CORNISH catalogue \citep{Purcell2013} or \citet{Walsh1998} are listed as {\uchii} regions. Parallel or already prior to this stage clumps appear as infrared-bright luminous sources due to the internally heated gas, often referred to as infrared bright massive clumps (cf.\,\citealp{Csengeri2016, Konig2017}). This stage is frequently characterised by rich molecular emission mainly originating from complex organic molecules leading to the emergence of hot cores. Many of our targets are well-characterised rich molecular hot cores that we also indicated in Table\,\ref{tab:sources}.
The two quiescent sources G23.21--0.3 and G34.41+0.2 lack bright emission at mid-infrared wavelengths and are likely to correspond to the earliest stages in the evolutionary sequence leading to high-mass star and cluster formation. For this sample of sources, this classification approach is in agreement with a classification based on the physical properties of the clumps, using their bolometric luminosity ($L_{\rm bol}$) and clump mass \citep[cf.][]{Konig2017}.
We list the sources and give a summary of the observed lines and transitions in Table\,\ref{tab:sources}. 

This paper is organised as follows. In Section\,\ref{sec:obs}, we introduce the  observed data and the used datasets, and in Section\,\ref{sec:results} we present our observational results. We perform a detailed non-local thermodynamic equilibrium (non-LTE) radiative transfer modelling in Section\,\ref{sec:ratran} and discuss our results in Section\,\ref{sec:discussion}.

%
\begin{table*}
\caption{List of target sources and observed species.}             
\label{tab:sources}      
\centering          
\begin{tabular}{l c r c c c c c c }
\hline\hline       
  Source  &  $\alpha$  & $\delta$~~~~~~~~~  & $\varv_{\rm lsr}$  & obs. species  & OD  & Type \\ 
    & (J2000) & (J2000)~~~ & (\kms) & & detection\\
\hline                    
G10.47+0.03        &  18:08:38.20 & $-$19:51:50.0 &       ${{+66.8}}^{{{{(a)}}}}$ & OD, OH~   & abs & {\uchii}, hot core{{${^{(1,2,3)}}$}} \\  
G23.21$-$0.3        & 18:34:54.91 &  $-08$:49:19.2      & $+76.8^{{{(a)}}}$ &  OD~  & abs & quiescent\\
G34.26+0.15         & 18:53:18.49 &  01:14:58.7    & ${+58.1}^{{{(a)}}}$ &  OD, OH, $^{18}$OH~   & abs & {\uchii}, hot core{{${^{(1,3)}}$}} \\
G34.41+0.2         & 18:53:18.13 &  01:25:23.7     & $+57.6^{{{(a)}}}$ &  OD~  & -- & quiescent  \\
G35.20$-$0.7        & 18:58:12.93 &  01:40:40.6   & $+33.4^{(a)}$ &  OD, OH$^\dagger$  & abs & hot core{{${^{(4)}}$}} \\
W28A (G5.89$-$0.4)         & 18:00:30.40 & $-$24:04:00.0  & ${+11.0}^{{{(a)}}}$ &  OD, OH~  & uncertain  & {\uchii}, hot core{{${^{(3)}}$}} \\
W31C (G10.62)         & 18:10:28.69 &  $-19$:55:50.0 & $-2.0^{{{(b)}}}$ &  OD, OH  & abs & {\uchii}, hot core{{${^{(1)}}$}} \\
W33A           & 18:14:39.52 &  $-17$:52:00.5   & $+36.{{7}}^{{{(a)}}}$ &  OD, OH$^\dagger$  & -- & {\uchii}, hot core{{${^{(5,6)}}$}} \\
W49N           & 19:10:13.20 &  09:06:12.0      & ${+12.8}^{{{(a)}}}$ &  OD, OH~  & two comp. abs  & {\uchii}, hot core{{${^{(1,3)}}$}} \\
W51e2          & 19:23:43.96 &  14:30:34.6   & $+58.{2}^{{{(a)}}}$ & OD  & two comp. abs & {\uchii}, hot core {{${^{(7)}}$}} \\
W51d           & 19:23:39.95 &  14:31:08.0  & $+58.{2}^{{{(a)}}}$ & OD~  & -- & \uchii{{${^{(1,7)}}$}}  \\
G327.29$-$0.6       & 15:53:08.55 & $-$54:37:05.1  & $-44.7^{(a)}$ & OD, OH  & abs & hot core {{${^{(7)}}$}}\\
G351.58$-$0.4       & 17:25:25.03 & $-$36:12:45.4  & $-95.9^{(a)}$ & OD, OH, $^{18}$OH  & abs & {\uchii}, hot core{{${^{(7,8)}}$}}\\
\hline                  
\end{tabular}
\tablefoot{Columns are, left to right, source designation, J2000 coordinates, LSR velocity, observed species, information on OD detection, and information on source type.
$^{(a)}$ Based on C$^{17}$O ($J$=3--2) measurements with the APEX telescope from \citet{Giannetti2014} and \citet{Wyrowski2016}.
$^{(b)}$ Based on NH$_3$ (1,1) measurements with the Effelsberg telescope from \citet{Wienen2012}.
$^{\dagger}$ Only the excited OH lines have been observed.
$^{(1)}$ \citet{Purcell2013}
$^{(2)}$ \citet{Rolffs2011}
$^{(3)}$ \citet{Hatchell1998}
$^{(4)}$ \citet{SM2013}
$^{(5)}$ \citet{vanderTak2005}
$^{(6)}$ \citet{Immer2014}
$^{(7)}$ \citet{Rolffs2011b}
$^{(8)}$ \citet{Walsh1998}
\\
}\end{table*}

   \begin{figure}
   \centering
\resizebox{\columnwidth}{!}{\includegraphics{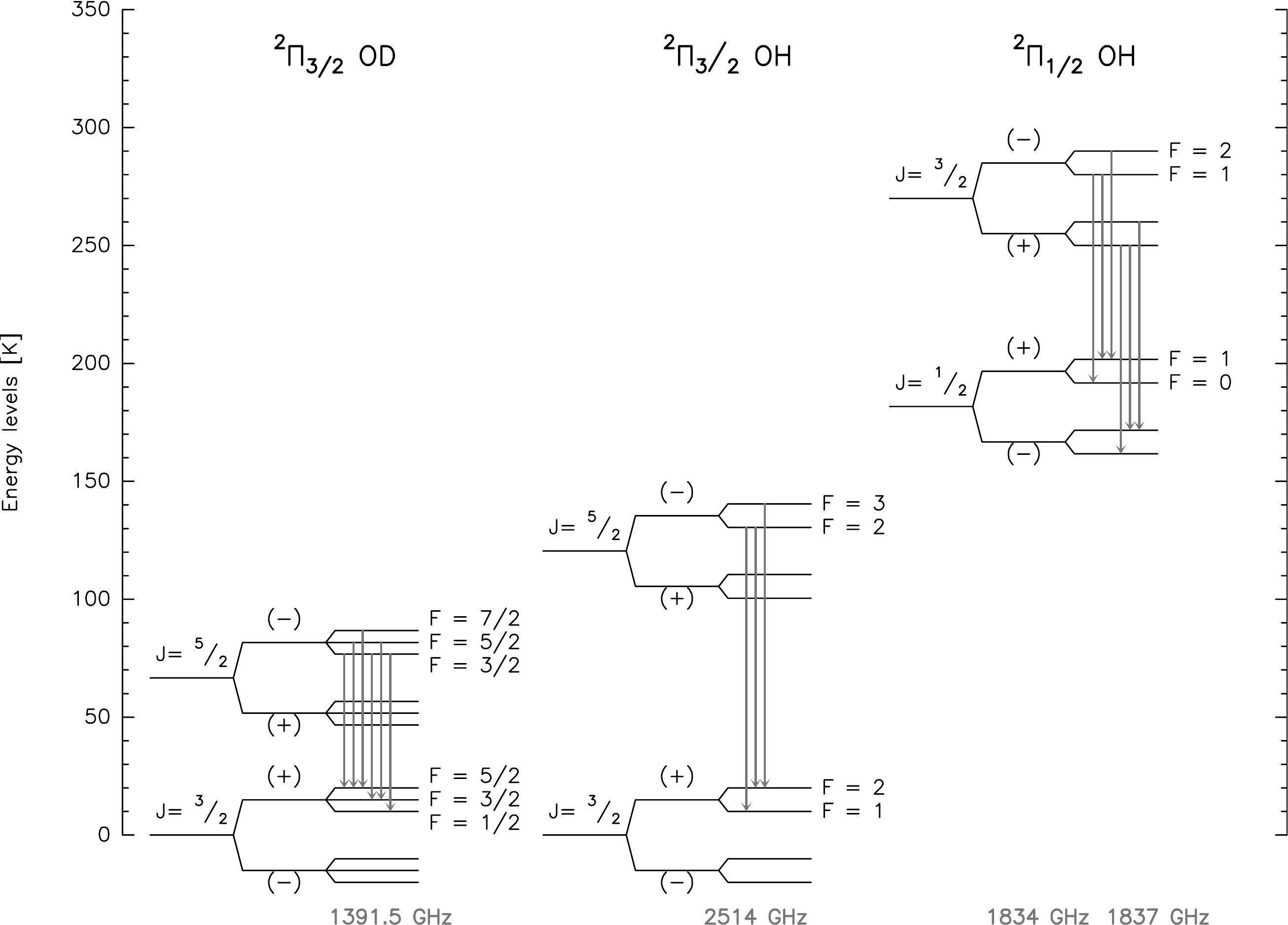}}
    \caption{Ground state and first excited energy levels of the \oht\ OD, OH, and \oho\ OH states. Grey arrows mark the transitions discussed in this study, whose frequencies are given at the bottom. The zero level of the energy scale refers to the vibrational ground states (2658 K and 1941 K for OH and OD, respectively).}
         \label{fig:energy}
   \end{figure}

\begin{table*}[ht!]
\setlength{\tabcolsep}{2pt}
\caption{List of molecules and their transitions. }
\label{tab:lines}      
\centering          
\begin{tabular}{l l l r c c c l c c c } 
\hline\hline       
Molecule       & Transition & Hyperfine & Frequency & \multicolumn{1}{c}{$\Delta\varv_{\rm HFS}{{^\star}}$} & $A_{\rm E}^\dagger$  &
$E_{\rm up}/k$~~~ & Instrument & $B_{\rm eff}$ & $HPBW$ & Reference \\ 
              &            & component & (GHz)~~~  & (\kms) & (s$^{-1}$)  & [K]~~~
    &      & & (\arcsec) & \\
\hline                    
OD     & $^2\Pi_{3/2}\,\, J=5/2\leftarrow 3/2$ & $F=3/2^-\leftarrow 5/2^+$ & 
        1391.4857 & $+1.9$ & 2.3($-$4) & 67 & SOFIA/GREAT & 0.67 & 20\arcsec & (1) \\
       &                 & $F=5/2^-\leftarrow 5/2^+$ &  1391.4895  &  $+1.1$ & 3.7($-$3)  &            \\
       &                 & $F=7/2^-\leftarrow 5/2^+$ &  1391.4947  &  $~~~0.0$ & 2.3($-$2)  &            \\
       &   & $F=3/2^- \leftarrow3/2^+$ & 
         1391.4980  &   $-0.7$ &  5.6($-$3)     &  \\
       &                                   & $F=5/2^- \leftarrow 3/2^+$ &
         1391.5017  &   $-1.5$ & 2.0($-$2) &             \\ 
       &                                   & $F=3/2^- \leftarrow 1/2^+$ &
         1391.5051 &  $-2.3$& 1.8($-$2) &       \\
OH       & $^2\Pi_{1/2}\,\, J=3/2\rightarrow 1/2$ & $F=1^- \rightarrow1^+$ & 
          1834.7350  &$+1.9$& 2.1($-$2) &   270    & SOFIA/GREAT & 0.67$^\ddagger$ &16\arcsec & (1) \\
       &                                   & $F=2^- \rightarrow 1^+$ &
         1834.7469  & $~~~0.0$  & 6.4($-$2)  &             \\ 
       &                                   & $F=1^- \rightarrow 0^+$ &
         1834.7499 & $-0.5$ & 4.2($-$2) &       \\
       & $^2\Pi_{1/2}\,\, J=3/2\rightarrow 1/2$ & $F=1^+ \rightarrow1^-$ & 
         1837.7461  & $+11.5$  & 2.1($-$2) &  270     & SOFIA/GREAT & 0.67$^\ddagger$ & 16\arcsec \\
       &                                   & $F=2^+ \rightarrow 1^-$ &
         1837.8163  & $~~~0.0$  & 6.4($-$2) &             \\ 
       &                                   & $F=1^+ \rightarrow 0^-$ &
         1837.8365 &  $-3.3$ & 4.3($-$2) &       \\
$^{18}$OH     & $^2\Pi_{3/2}\,\, J=5/2\leftarrow 3/2$ & $F=2^- \leftarrow 2^+$ & 
         2494.6809  & $+1.7$ & 1.4($-$2) &  120      & SOFIA/GREAT & 0.70 & 12\arcsec  & (1) \\
       &                                   & $F=3^- \leftarrow 2^+$ &
         2494.6950  & $~~~0.0$  & 1.4($-$1) &             \\ 
       &                                   & $F=2^- \leftarrow 1^+$ &
         2494.7342 & $-4.7$ & 1.2($-$1) &       \\        
 \hline
OH     & $^2\Pi_{3/2}\,\, J=5/2\leftarrow 3/2$ & $F=2^- \leftarrow 2^+$ & 
         2514.2987  & $+2.1$ & 1.4($-$2) &  121      & SOFIA/GREAT & & 12\arcsec  & (2) \\
       &                                   & $F=3^- \leftarrow 2^+$ &
         2514.3167  & $~~~0.0$  & 1.4($-$1) &             \\ 
       &                                   & $F=2^- \leftarrow 1^+$ &
         2514.3532 & $-4.4$ & 1.2($-$1) &       \\        
 HDO & $1_{1, 1}-0_{0, 0}$ & & 893.639~~ & & 8.4($-$3)& 43 & \textsl{Herschel}/HIFI & & 24\rlap{.}{\arcsec}1 & (3) \\
  & &   & & & && APEX/CHAMP & & 7\arcsec & (4) \\
 p-H$_2^{18}$O & $1_{1, 1}-0_{0, 0}$ & & 1101.698~~  & & 1.8($-$2) & 53 & \textsl{Herschel}/HIFI & & 19\rlap{.}{\arcsec}2 & (5) \\
 \hline                  
\end{tabular}
\tablefoot{Frequencies, Einstein coefficients, and upper-level energies are
from the JPL \citep{1998JQSRT..60..883P} database. The $^{18}$OH line is only available for G34.26+0.15. 
HDO and the H$_2^{18}$O line has only been used for G34.26+0.15. Molecules below the horizontal line are archival data.
${{^\star}}${Velocity offset of the hfs components.}
$^\dagger$ The Einstein coefficients are listed in the form of $A(B)=A\times10^B$. $^\ddagger$ This line has been observed in Cycle~4 towards the source G35.20$-$0.7 with a main beam efficiency of 0.70. (1) This work, (2) \citet{Wiesemeyer2016}, (3) \citet{Coutens2014}, (4) \citet{Liu_thesis}, (5) \citet{Flagey2013}.}
\end{table*}

\section{Observations and data reduction}\label{sec:obs}

\subsection{SOFIA/GREAT observations}
We used the high spectral resolution GREAT receiver\footnote{GREAT is a development by the MPI für Radioastronomie
and KOSMA/Universität zu Köln, in cooperation with the MPI
für Sonnensystemforschung and the DLR Institut für Optische
Sensorsysteme.} \citep{Heyminck2012, Risacher2016}
on board the Stratospheric Observatory for Infrared Astronomy (SOFIA) \citep{Young2012} to observe rotational transitions of the OD and OH lines.
Our primary focus was the $^2\Pi_{3/2}$ ground-state ($J = 5/2 - 3/2$) transition of 
OD at 1391.5\,GHz, which was complemented towards the majority of our sample by observations of the $^2\Pi_{{1/2}}$ excited-state ($J  = {3/2} - {1/2}$) $\Lambda$-doublet  transitions of OH at 1834.7 (163.4~$\mu$m) and 1837.8~GHz (163.1~$\mu$m), respectively. 

The observations were carried out as part of GREAT consortium guaranteed observing time in the observatory's Cycle~1, and Cycle~4 programmes. Best spectroscopic baselines were obtained by means of the chop-and-nod procedure (double-beam switching), with chop amplitudes of 30\arcsec\, to 60\arcsec, operated at a frequency of 1~Hz. Owing to the double-sideband reception of GREAT, the selection of the sideband and intermediate frequency was optimized with regard to the atmospheric transmission (so as to avoid ozone features) and to receiver sensitivity. 
In Cycle~1 for OD, the signal was recorded at 
1391.5~GHz using the L1 band of GREAT tuned in the upper side-band, and for the $^2\Pi_{{1/2}}$ excited-state OH line
using the L2 band at 1837.8~GHz tuned in the upper side-band. 
The 2.5\,GHz wide intermediate frequency band was tuned to 1.7\,GHz offset, allowing us to  simultaneously cover the 1834.7 GHz $\Lambda$-doublet pair in the image band. In Cycle~4, the $^2\Pi_{{1/2}}$ excited-state OH line was observed using the GREAT low frequency array module (LFA) array and was also tuned to 1837.8~GHz in the upper side-band. 
{We also recorded the $^2\Pi_{{3/2}}$ ground-state $^{18}$OH line at 2494.7~GHz towards G34.26+0.15 and G351.58$-$0.4 using the M$_{\rm a}$ channel of GREAT {\citep{Heyminck2012}} in Cycle~1 in July 2013. The receiver was tuned to 2494.420\,GHz in the upper side band. 

We used the extended fast Fourier transform spectrometers (XFFTS) as backends with 2.5\,GHz bandwidth and 76 kHz spectral resolution  in Cycle~1, and the FFTS4G in Cycle~4 providing a 4\,GHz bandwidth.
The median single-sideband system temperature was $\sim$1200\,K at 1391.5~GHz, and $\sim$1600\,K at 1834.7\,GHz (Rayleigh-Jeans equivalent antenna temperatures at systemic velocity).
The pointing was established with the optical guide cameras providing an accuracy 
of 2-3{\arcsec},  {and linked to the instrument reference frame by peaking up on Mars or Jupiter right after each reinstallation of the GREAT receiver system}. The observed lines for each source are listed in Table\,\ref{tab:sources}, and a detailed summary of the observed transitions is listed in Table\,\ref{tab:lines}.

The data were calibrated using standard procedures based on the KOSMA/GREAT calibrator \citep{Guan2012}, and for the data analysis  
we used the GILDAS\footnote{See http://www.iram.fr/IRAMFR/GILDAS/} software. The transmission correction was based on the AM atmosphere model\footnote{S. Paine, SMA Memo \#152, Smithsonian Astrophysical Observatory}, scaling the measured total power by a forward efficiency of 97\%. We fine-tuned the correction for the ozone features by scaling their column densities in AM by factors of order unity until they disappeared against the continuum signal. The data was then converted to a Rayleigh-Jeans equivalent main-beam brightness temperature ($T_{\rm mb,RJ}$) scale using a beam efficiency of 67\% at both frequencies for Cycle 1, and of 70\% for the 1837.8\,GHz
observations in Cycle 4. {For the $^{18}$OH line, we used  {a beam efficiency of 70\%.}

For the OD line, we obtained the continuum emission by means of a dedicated double-sideband calibration. The $\Lambda$-doublet pair of the $^2\Pi_{{1/2}}$ excited-state OH lines originates from different sidebands and was therefore calibrated considering the different atmospheric transmission in the image and signal bands. The spectra from the signal and image band calibration were then stitched together at an emission-free velocity to obtain a composite spectrum with the respective atmospheric calibration for the spectral features from both the signal and image band.

For both the \oht\ OD ($J$=5/2--3/2) and the \oht\ OH ($J$=3/2--1/2) lines, the spectra were first summed up to define windows 
excluding the lines for the baseline fitting. Then we used 
a polynomial baseline to fit the continuum emission and subtracted the fitted continuum from each scan 
individually. The spectra were then averaged with noise weighting and 
were smoothed to $\sim$ 0.65 and 0.75 \kms\ at 1391.5 and 1837.8\,GHz, respectively. 
We reach an average noise of 84 and 87 mK for these velocity resolutions. 
The half-power beam width ($HPBW$) is $\sim$20{\arcsec} at 1391.5\,GHz, 16{\arcsec} at 1837.8~GHz.
The same procedure has been applied for the $^{18}$OH line; however, the baseline  variations needed a more careful treatment for G34.26+0.15 due to instrumental effects (see App.\,\ref{app:18oh}).

\subsection{Archival SOFIA data}

To put better constraints on the origin of the emission, we used data resulting from SOFIA observations of the $^2\Pi_{3/2}, J = 5/2 - 3/2$
OH rotational ground-state transition  at 2514~GHz from \citet{Wiesemeyer2012,Wiesemeyer2016}.
The observational setup and data reduction steps are similar to those described above. The final spectra were smoothed to 0.9~\kms\
velocity resolution and have a median noise level of $\sim$0.3~K. The $HPBW$  is 12{\arcsec} at 2514\,GHz. 

\subsection{Archival \textsl{Herschel}  data}
To compare the OD and HDO line profiles, we used spectra of the HDO ($J_{K,A}=1_{1, 1}-0_{0, 0}$) line at 893.6\,GHz \citep{Coutens2014} observed by the \textsl{Herschel}/HIFI instrument, where available. In addition to this, to assist the detailed modelling of G34.26+0.15, we also used a spectrum of the p-H$_2^{18}$O ($1_{1, 1}-0_{0, 0}$) line at 1101.698\,GHz  \citep{Flagey2013}.  These observations were made in the framework of the Probing Interstellar Molecules with Absorption Line Studies (PRISMAS) Guaranteed Time Key programme \citep{Gerin2010, Neufeld2010}. 
PRISMAS observations employed a dual-beam-switching pointed observing mode with a reference position of $\sim$3\arcmin\  from the source. We obtained the level 2.5 data products from the Herschel Science Archive and processed the data with the IRAM/GILDAS software. We averaged the horizontal and vertical polarisations and summed up the spectra from the three different local oscillator (LO) tunings that have been used to identify line contamination from the image band. We fitted the spectral baseline over a narrow velocity range around the rest velocities ($\varv_{\rm lsr}$) of the sources and assumed a sideband gain ratio of unity to correct for the double-sideband receiver. We then used a beam efficiency of 74\%  \citep{Roelfsema2012} to convert the spectra to a main beam brightness temperature scale.

For G327.29$-$0.6, we used the HDO ($J_{K,A}=1_{1, 1}-0_{0, 0}$) line at 893.6\,GHz from the WISH \citep{vanDishoeck2021} programme also observed with the \textsl{Herschel}/HIFI instrument. The data were processed in the same way as above, with the exception that in this case a single frequency setup was used to cover the line.

\subsection{Archival APEX data}

Towards the source G351.58$-$0.4, the complementary data for the HDO ($J_{K,A}=1_{1, 1}-0_{0, 0}$) line at 893.6\,GHz   were observed with the Atacama Pathfinder Experiment Telescope (APEX, \citealp{Gusten2006}) using the CHAMP+ receiver (\citealp{Liu_thesis}  PhD thesis).

   \begin{figure}[!ht]
   \centering
   \includegraphics[width=\hsize]{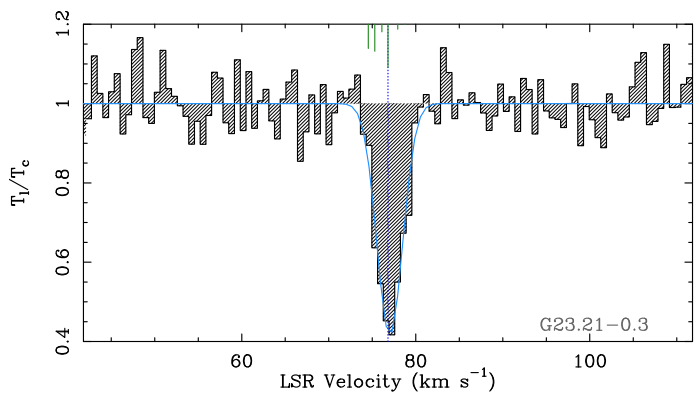}
    \caption{Example of the OD $^2\Pi_{3/2}$ ground-state ($J = 5/2 - 3/2$) absorption towards the quiescent source G23.21-0.3. The spectra are normalised to the (single side-band) continuum level, and thus show the line to continuum ratio. The rest frequency is set to the strongest hyperfine transition, $F$=7/2$^-$--5/2$^+$, at 1391494.70\,MHz. The blue  dotted line marks the systemic LSR velocity of the source, and the grey dotted line shows the value of 1. The blue line shows a single component Gaussian fit to the spectrum. The positions and the relative strength of the hyperfine components are shown in green. }
    \label{fig:OD_G2321}
     \end{figure}

   \begin{sidewaysfigure*}[!]
   \centering
   \includegraphics[width=\hsize]{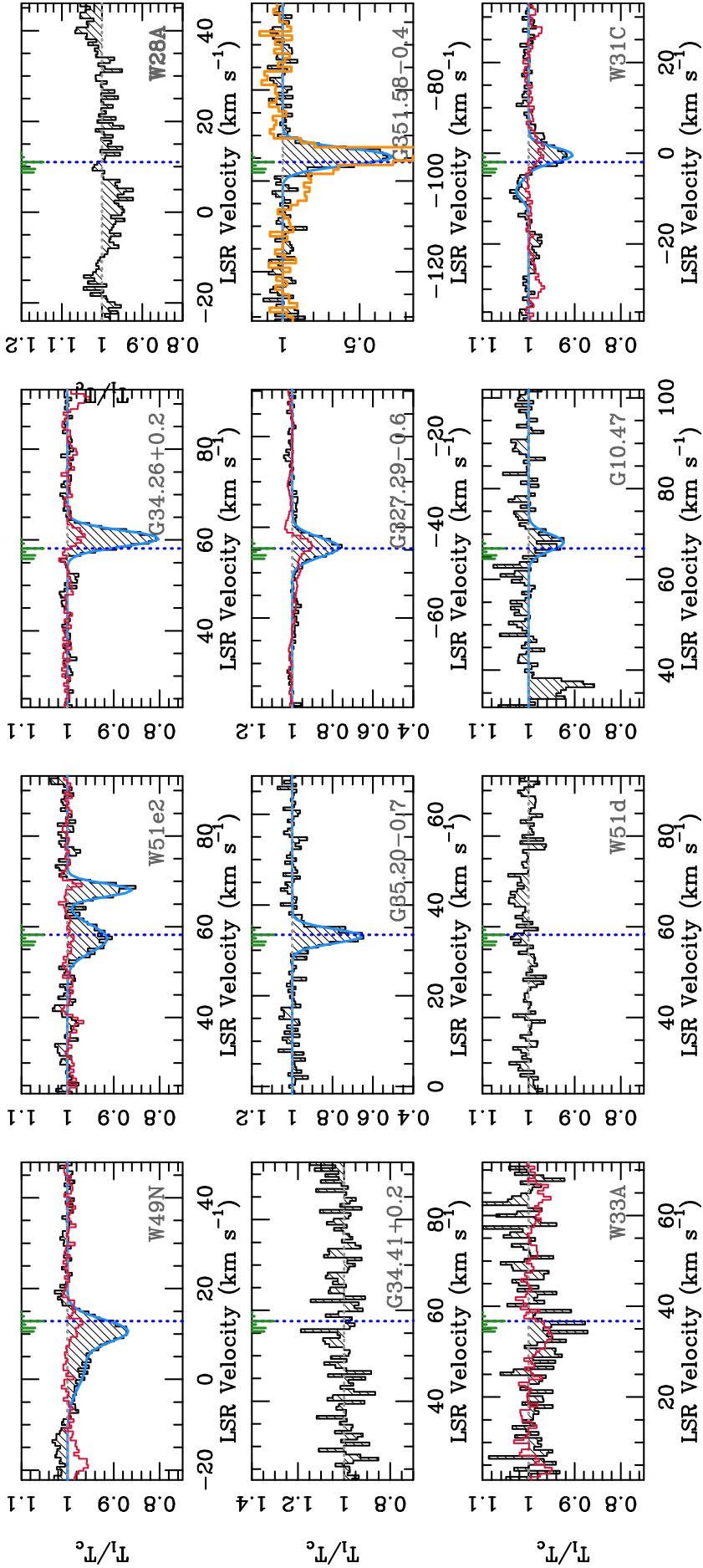}
    \caption{Overview of the OD $^2\Pi_{3/2}$ ground-state ($J = 5/2 - 3/2$) transition towards the sample. The spectra are normalised to the continuum level and thus show the line-to-continuum ratio. The rest frequency is set to the strongest hyperfine transition, $F$=7/2$^- - 5/2^+$, at 1391494.70\,MHz. The vertical blue  dotted line marks the systemic LSR velocity  of the source (Table\,\ref{tab:sources}),  horizontal grey dotted line shows the  intensity ratio of 1. The light blue line shows a single or a two-component Gaussian fit to the spectrum. The positions and the relative strength of the hyperfine components are shown in green. The red spectra correspond to the HDO ($J_{K,A}=1_{1, 1}-0_{0, 0}$) line at 893.6\,GHz from \textsl{Herschel}/HIFI, the orange spectrum shows the same line observed with the APEX telescope.  The source name is labelled in the lower right corner of each panel.}
         \label{fig:od_all}
   \end{sidewaysfigure*}

   \begin{figure*}[!]
   \centering
   \includegraphics[width=0.42\hsize]{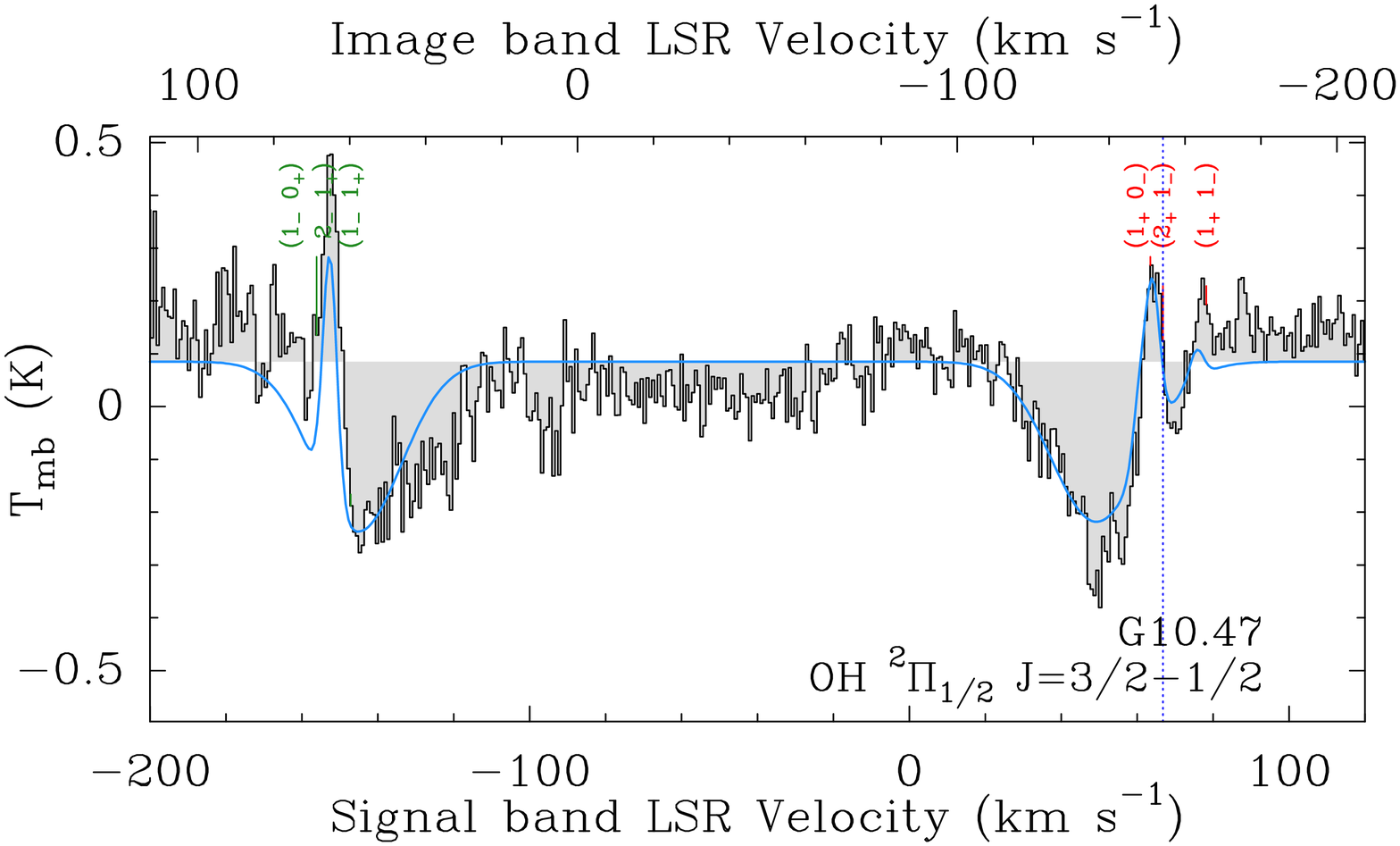}
    \includegraphics[width=0.42\hsize]{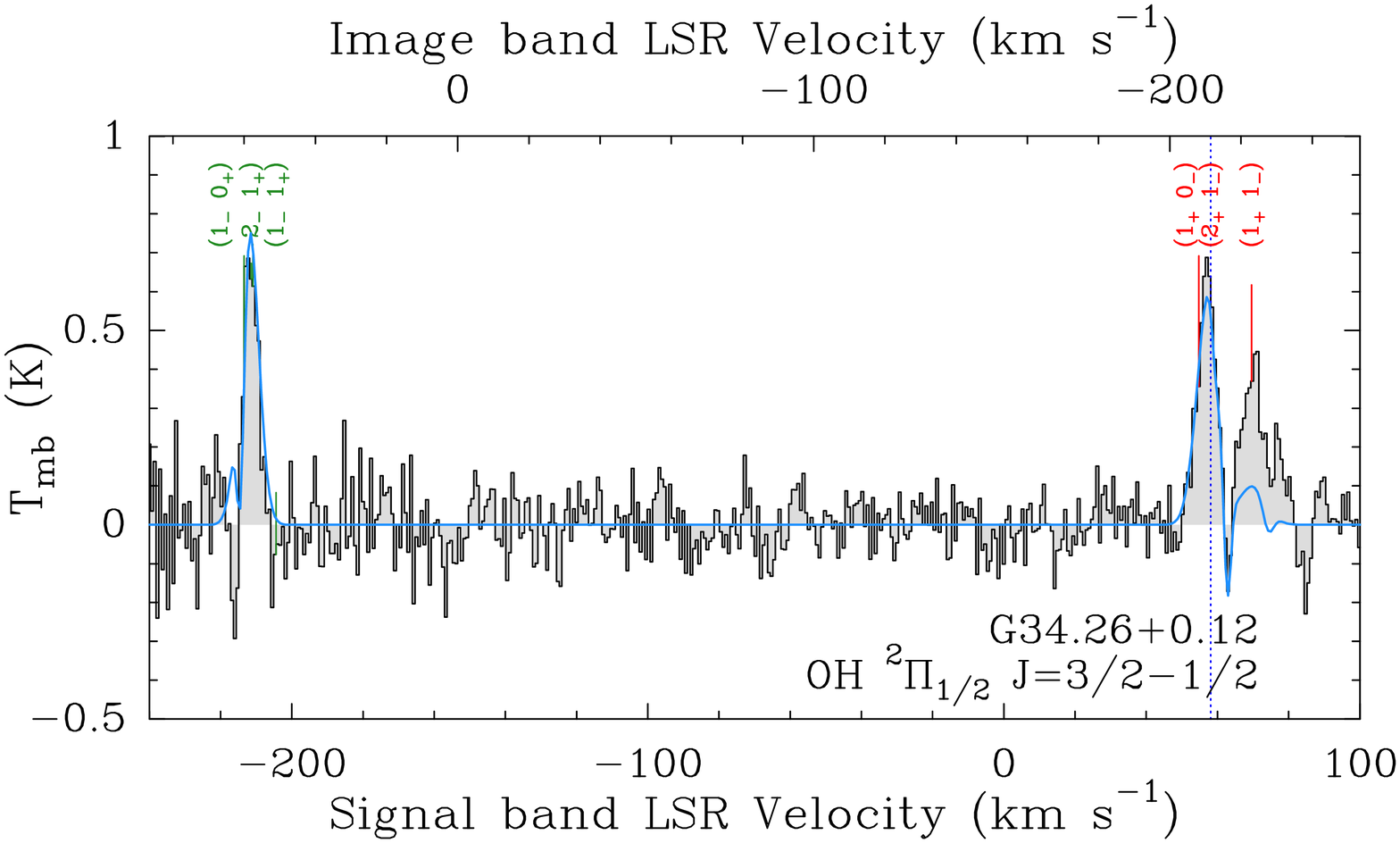}
      \includegraphics[width=0.42\hsize]{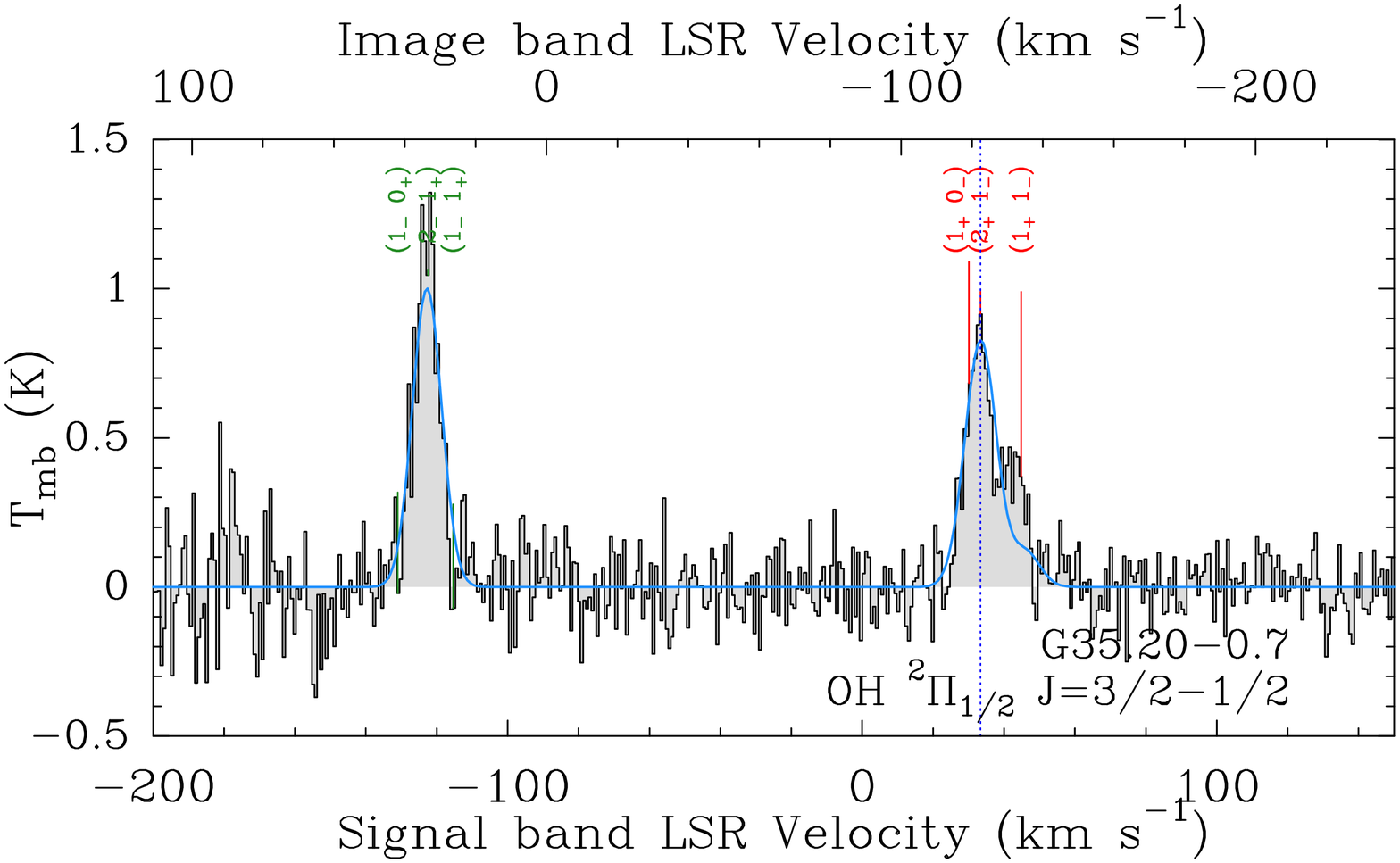}
   \includegraphics[width=0.42\hsize]{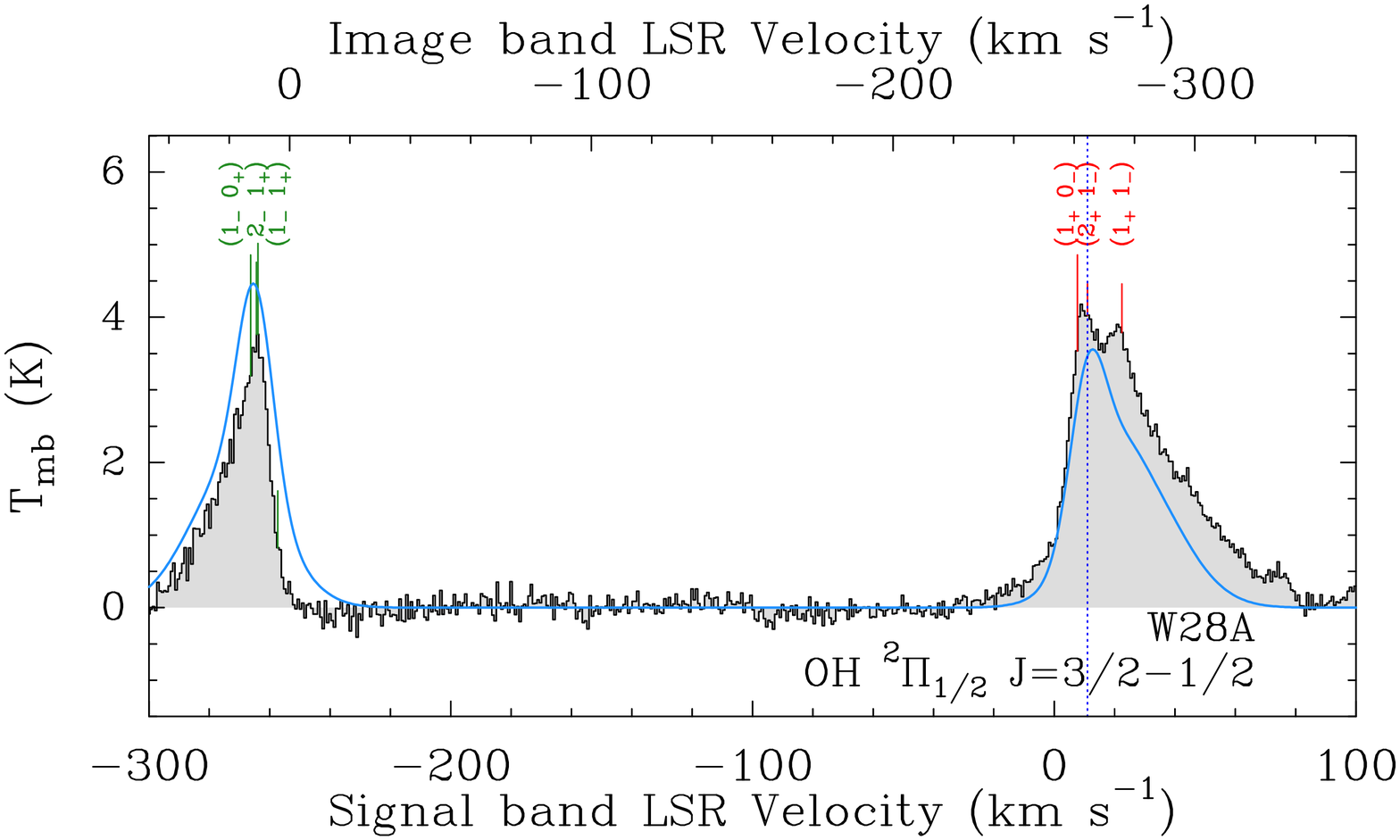}
  \includegraphics[width=0.42\hsize]{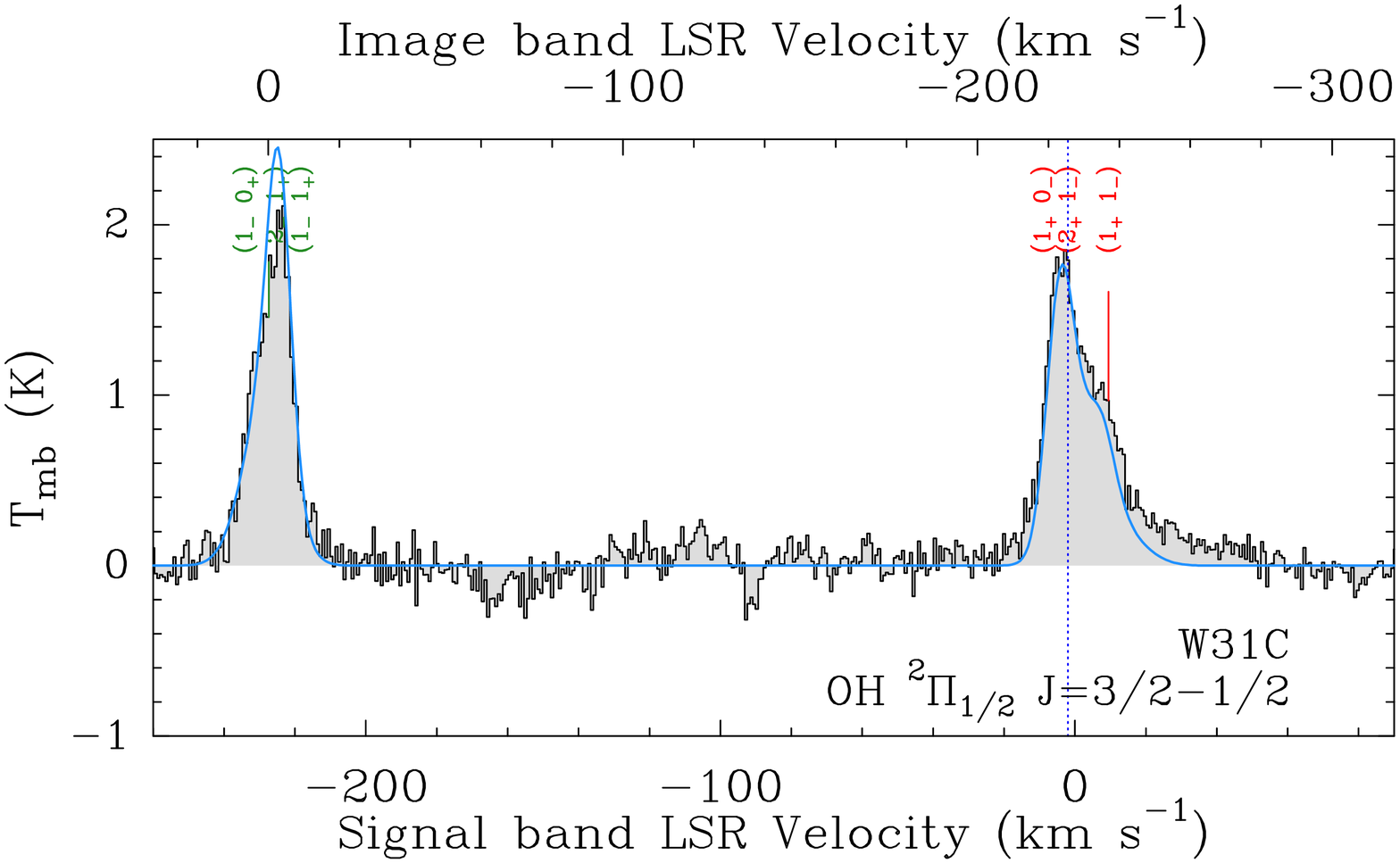}
  \includegraphics[width=0.42\hsize]{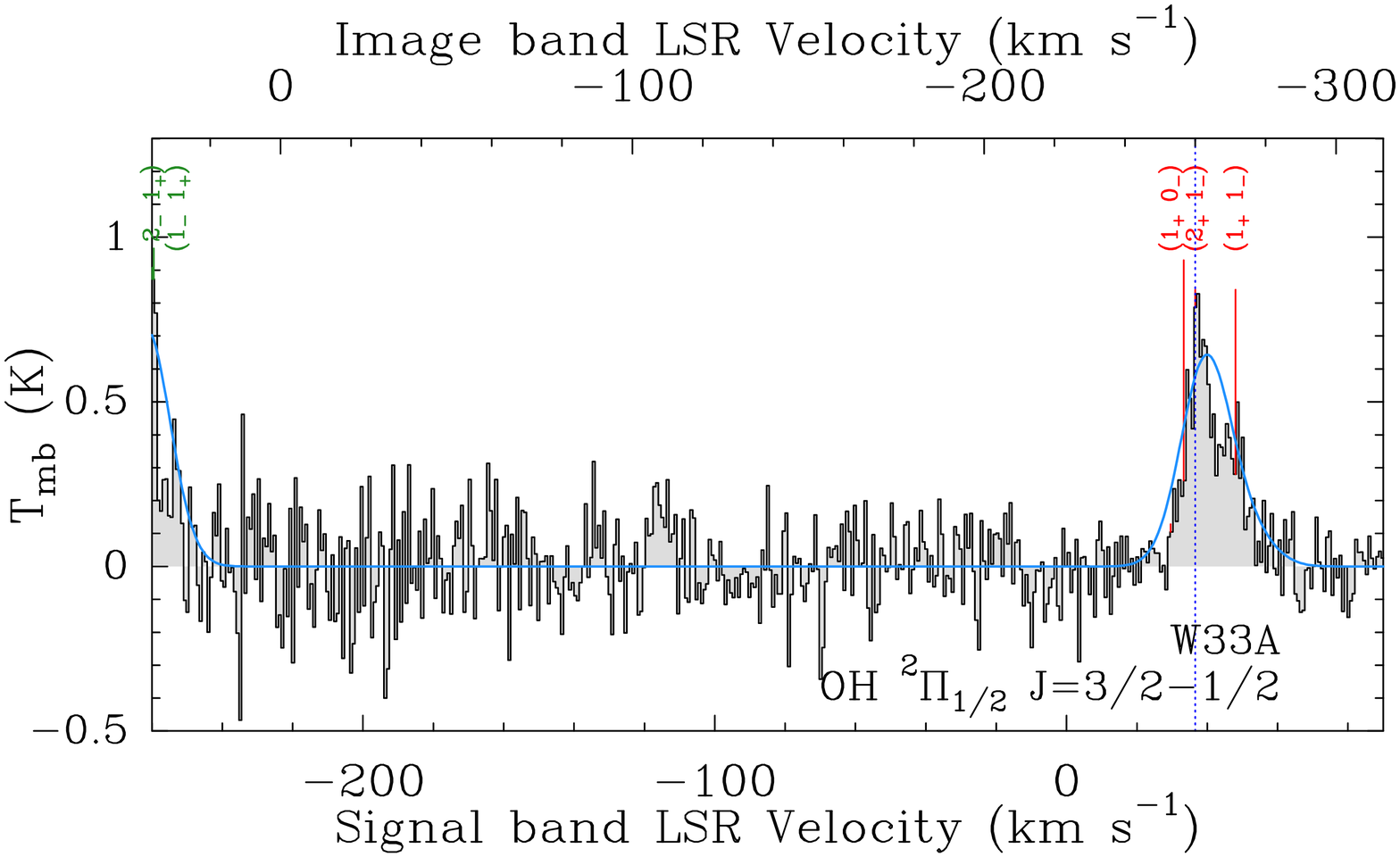}
  \includegraphics[width=0.42\hsize]{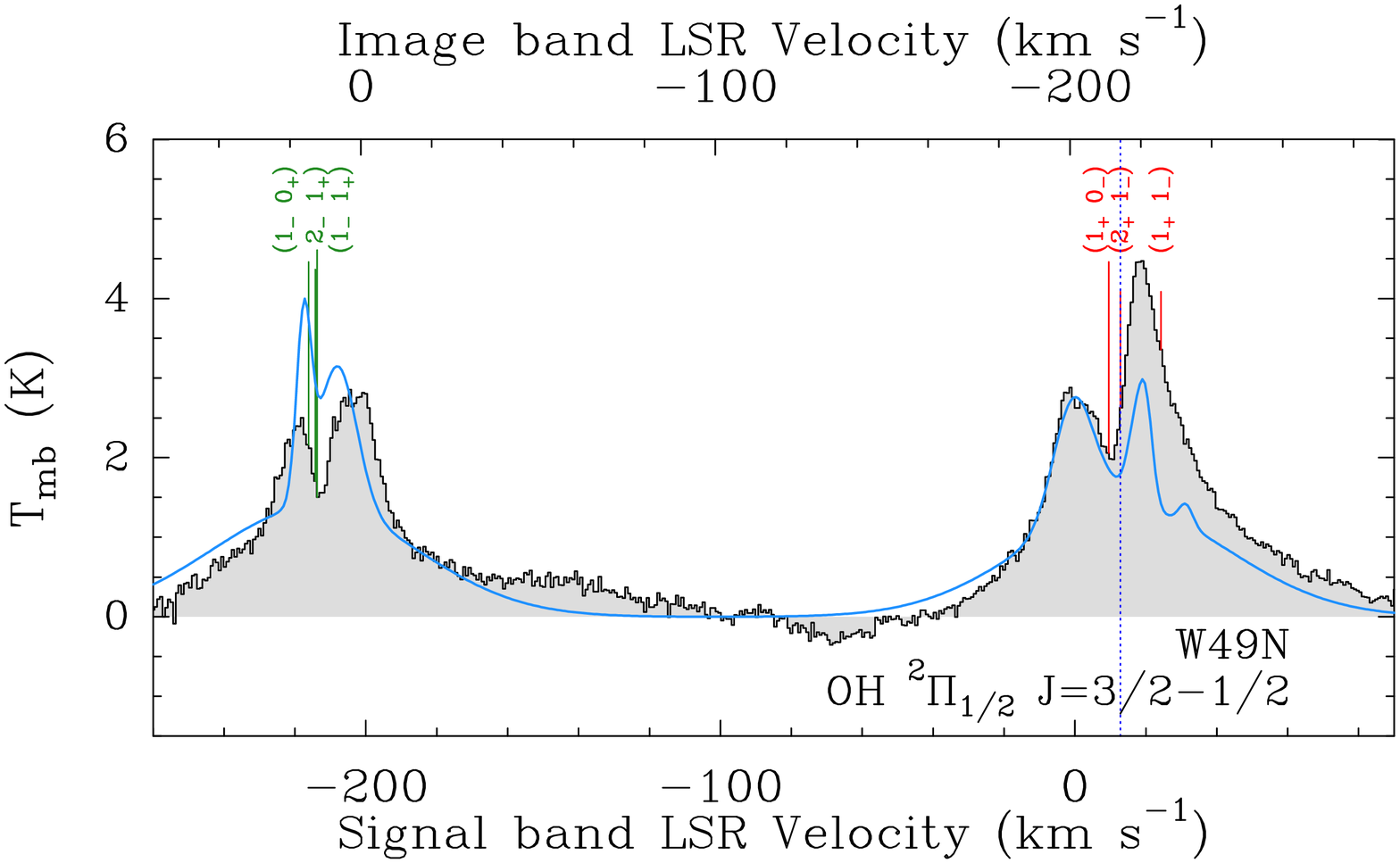}
 \includegraphics[width=0.42\hsize]{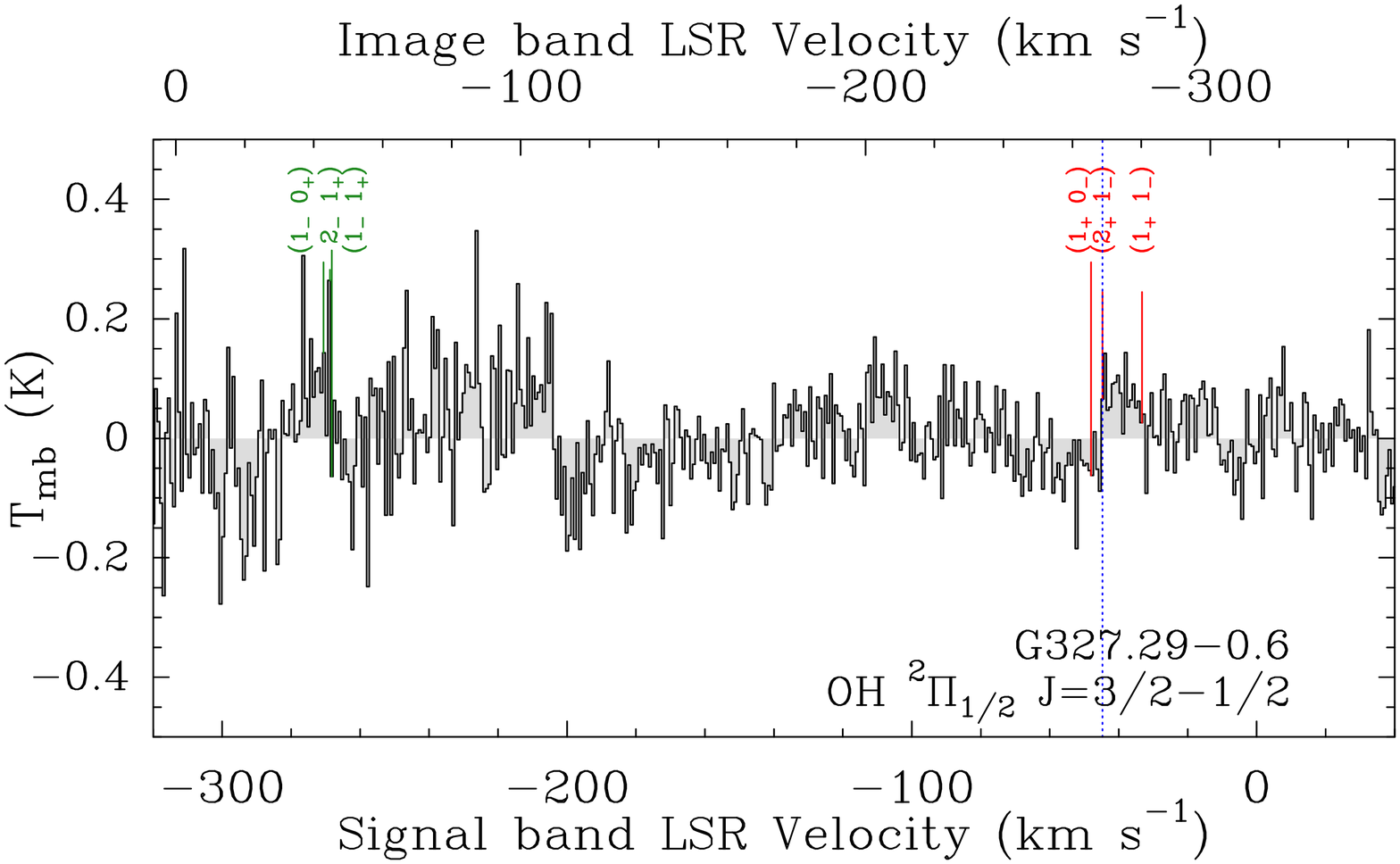}
  \includegraphics[width=0.42\hsize]{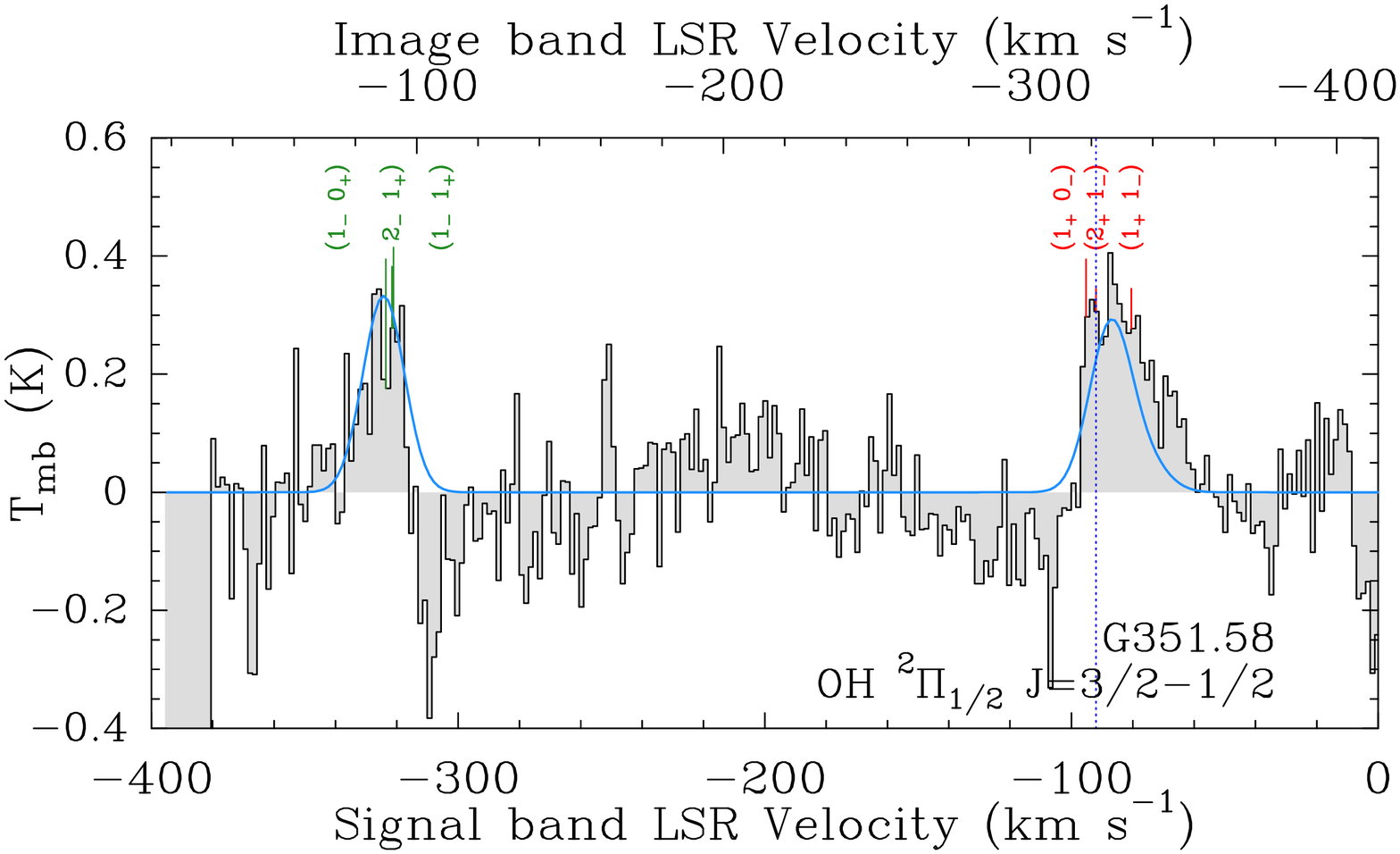}

    \caption{Overview of OH $^2\Pi_{1/2}$ ($J = 3/2 - 1/2$) transition towards the sample. The blue dotted line marks the systemic LSR velocity of the source in the signal band. The positions of the hfs lines are labelled in red in the signal and in green in the image band. The HFS fit to the spectra is shown in light blue.  {The velocity resolution is 0.75\,\kms\ for all sources, except for G351.58$-$0.4, where we show the spectrum at a velocity resolution of 1.5\,\kms}.
    }
         \label{fig:oh_all}
  \end{figure*}

\section{Results}\label{sec:results}

\begin{table*}[ht!]
\caption{Fit results and gas parameters estimated from the $^2\Pi_{3/2}$ OD $J$=5/2--3/2 line. }
\label{tab:results-od}
\begin{tabular}{l c c c c c c c c c}
\hline\hline 
   &   $T_{\rm c}^{\rm OD}$$^{(a)}$ & $\sigma_{\rm rms}$ & method$^{(b)}$ & $T_{\rm mb}^{(c)}$  & $\varv_{\rm lsr}^{(d)}$ & $\Delta \varv_{\rm FWHM}^{(c)}$ & $T_{\rm l}/T_{\rm c}^{(e)}$ & $\tau_{\rm tot}^{(e)}$ & $N_{\rm tot}^{(e)}$ \\
   & $$(K) & (K) & & (K) & (\kms) & (\kms) & (K) & & $\times10^{12}$ (cm$^{-2}$)\\
\hline
G10.47+0.03   & 3.7 & 0.17 & Gaussian   & $-0.27\pm0.06$   & $68.13\pm0.37$ & $4.45\pm0.69$ &  0.07$\pm$ 0.05 &  0.08$\pm$ 0.04 & 5.30$\pm$ 2.65 \\
               &      &            & hfs   &  & $68.90\pm0.37$ & $3.63\pm0.85$ &  & 0.08 \\
G23.21-0.3  &  0.8    & 0.06 & Gaussian   & $-0.47\pm0.05$ & $77.01\pm0.13$ & $3.49\pm0.27$ & 0.59$\pm$ 0.09 &  0.89$\pm$ 0.05  & 35.76$\pm$ 2.01\\
                   &           & &  hfs   &   & $77.80\pm0.13$ & $2.20\pm0.30$ & & 0.89\\                 
G34.26+0.15  &  6.5   & 0.09 & Gaussian   &  $ -1.31\pm0.05$& $60.31\pm0.05$ & $ 4.18\pm0.12$ & 0.20$\pm$ 0.01 & 0.23$\pm$ 0.01 & 13.61$\pm$ 0.59 \\ 
                    &           &           & hfs   &   & $61.10\pm0.05$ & $3.24\pm0.14$ & & 0.23    \\
G34.41+0.2   &  1.4  & 0.10 & -- & -- & -- & -- & -- & -- \\
G35.20$-$0.7  &  1.7 & 0.07 & Gaussian   & $-0.60\pm0.06$  & $33.13\pm0.11$ & $3.70\pm0.29$ &  0.35$\pm$ 0.04  & 0.44$\pm$ 0.03  & 20.89$\pm$ 1.42 \\
                    &        & &  hfs   &                                          & $33.90\pm0.12$ & $2.60\pm0.36$ & & 0.44     \\                  
W28A      & 3.9   & 0.11& -- & -- & -- & -- & -- \\

W31C   &   6.5 & 0.07 & Gaussian   & ~~$0.19\pm0.06$   &   $-8.29\pm0.51$ & $4.96\pm1.23$  &  --   & -- & -- \\
           &     &  &                            &  $-0.61\pm0.16$   &  $-0.39\pm0.17$ & $3.83^\dagger$  & 0.10$\pm$ 0.01 & 0.10$\pm$ 0.01 & 12.09$\pm$ 1.21\\

             &           &  & hfs   &    & $-2.65\pm0.66$  &  $9.10\pm2.19$ & & \\
             &           &  &       &    & $-0.30\pm0.66$  &  $6.62\pm2.19$ & & 0.1 \\

W33A    &  1.7  & 0.06  & -- & -- & -- & -- & -- \\
W49N   & 9.5   & 0.10 & Gaussian   &    $-1.13\pm0.09$ & $10.78\pm0.17$ &  $5.79\pm0.34$ &   0.12$\pm$ 0.01  & 0.13$\pm$ 0.01  & 27.78$\pm$ 2.14 \\
             &         &            &                     &   $-0.38\pm0.04$ & $3.30\pm0.83$   &  $12.26\pm1.06$ &  0.04$\pm$ 0.01 & 0.04$\pm$ 0.01 & 3.86$\pm$ 0.96\\             
            & 9.5   & & hfs   &     &  $3.97\pm1.20$ &  $11.70\pm1.52$  &  & 0.13 \\
             &         & &      &    & $11.50\pm0.18$ &  $5.28\pm0.45$  &  & 0.04 \\

W51e2 & 7.6 & 0.13 & Gaussian   &  $-1.06\pm0.13$& $68.06\pm0.14$ &   $3.63\pm0.30$ &  0.14$\pm$ 0.02&  0.15$\pm$ 0.02 & 4.47$\pm$ 0.60 \\
            &       &            &                   & $-0.66\pm0.09$ & $57.40\pm0.31$ &   $7.10\pm0.74$ & 0.09$\pm$ 0.02 &  0.09$\pm$ 0.02 &11.08$\pm$ 2.46 \\
            &       & & hfs   &   & $68.80\pm0.11$&   $1.63\pm0.32$ &  & 0.15 \\
            &       &  & &   & $58.2\pm0.31$ &   $6.74\pm0.81$ & & 0.09\\

W51d   & 4.7 & 0.08  & -- & -- & -- & -- & -- & --  \\
G327.29$-$0.6 &  5.6 & 0.08 & Gaussian  & $-1.32\pm0.08$& $-44.40\pm0.09$  & $5.22\pm0.23$ &  0.24$\pm$ 0.01  & 0.27$\pm$ 0.01 & 21.85$\pm$ 0.81\\
              &        &            & hfs  &  & $-43.60\pm0.9$  & $4.43\pm0.25$ & & 0.27   \\
G351.58$-$0.4 & 1.7   & 0.11 & Gaussian  & $-1.23\pm0.09$& $-94.69\pm0.11$  & $4.36\pm0.25$ & 0.72$\pm$ 0.08 & 1.29$\pm$ 0.05  & 69.27$\pm$ 2.68\\
               &      &           &hfs  & & $-93.90\pm0.11$  & $2.94\pm0.27$ & & 1.29 \\

\hline
\end{tabular}
\tablefoot{{$^{(a)}$} Main-beam brightness temperature of the single-sideband continuum at 1.391~THz (Rayleigh-Jeans scale).  {{$^{(b)}$} The fit corresponds to a one- or two-component Gaussian fit to the spectra, while the hfs method takes into account the relative strengths and velocity pattern of the hyperfine structure lines.}{$^{(c)}$} The errors are propagated from the HFS fit errors from CLASS using the respective formulae. {$^{(d)}$} The errors are constrained by the fitting procedure in CLASS, assuming a normal distribution of uncertainties. {$^{(e)}$} The errors are propagated from the fit errors and the noise of the data ($\sigma_{\rm rms}$) using standard formulation. $^\dagger$ The line width is fixed to that of the HDO line shown in Fig.\,\ref{fig:od_all}.\\}
\end{table*}

\subsection{The OD $^2\Pi_{3/2}$ ($J = 5/2 - 3/2$) transition }\label{sec:od_ground_state}
\subsubsection{Detection rates}
As an example, in Fig.\,\ref{fig:OD_G2321}  we show the line-to-continuum ratio spectrum ($T_{\rm l}/T_{\rm c}$) of the OD $^2\Pi_{3/2}$ ground-state ($J = 5/2 - 3/2$) transition at 1391.5\,GHz for the quiescent source G23.21-0.3. Spectra for the  rest of our sources  are shown in Fig.\,\ref{fig:od_all}. The lines typically exhibit a Gaussian profile that is a blend of the six hfs components of the $\Lambda$-doublet transition spanning a velocity range of $\sim$4.2\,\kms\ (see also Table\,\ref{tab:lines}).

The OD line is detected towards nine of the 13 targets (69\%) on the $3.5-22\sigma$ level. In many cases, it appears as a strong and narrow absorption feature. 
We first fitted the spectra with a single or a two-component Gaussian to extract the line parameters and evaluate the detection rate. The fitted line parameters are summarised in Table\,\ref{tab:results-od}, which also includes the measured continuum emission near the OD line frequency accounting for the double-sideband reception of the signal. Among the non-detections, some  sources may show 
a 
weak  feature close to our signal-to-noise criterion; for example, towards W33A an absorption feature around 2$\sigma_{\rm rms}$ is observed close to the expected LSR velocity ($\varv_{\rm lsr}$). 
In addition, towards W28A the features around the expected $\varv_{\rm lsr}$ could also be consistent with a weak and broad absorption component and also a line in emission 
at the 2$\sigma_{\rm rms}$ level. The line profiles of these components would  resemble those of the \oht\ OH line towards this source analysed in \citet{Leurini2015}, although we caution that in our data the baseline quality for 
this source is inferior to that of 
the other sources.

\subsubsection{The velocity components of OD absorption}

In Figs.\,\ref{fig:OD_G2321} and \ref{fig:od_all}, we show the  $\varv_{\rm lsr}$ of the sources from Table\,\ref{tab:sources}. The OD $^2\Pi_{3/2}$ ($J = 5/2 - 3/2$) absorption is typically redshifted compared to the source's $\varv_{\rm lsr}$, which is clearly the case for G34.26+0.15, G351.58--0.4, G10.47+0.03 and W31C, corresponding to 44\% of the detections. For these sources, the HDO ($J_{\rm K,A}$=$1_{1, 1}-0_{0, 0}$) line is also seen similarly redshifted, as well as the NH$_3$ ($J=3_{2+}-2_{2-}$) transition from \citet{Wyrowski2016}. The rest velocity at the peak of the absorption is on average $\sim$1.5\,\kms redshifted for both the OD $^2\Pi_{3/2}$ ($J = 5/2 - 3/2$) and NH$_3$ ($J=3_{2+}-2_{2-}$) lines.

The OD $^2\Pi_{3/2}$ ($J = 5/2 - 3/2$) absorption feature coincides well with the source systemic velocity for G35.20--0.7, W51e2 and G327.29--0.6, while the NH$_3$ ($J=3_{2+}-2_{2-}$) line is detected as redshifted by \citet{Wyrowski2016}. Considering the errors of the fit, which are larger by several factors for the OD line compared to the NH$_3$ line, we find that the $\varv_{\rm lsr}$ of OD and NH$_3$ are also consistent for these sources.

The OD $^2\Pi_{3/2}$ ($J = 5/2 - 3/2$) absorption is seen blueshifted as compared to the source's $\varv_{\rm lsr}$ towards W49N, often interpreted as a P-Cygni profile due to expanding gas. The same kinematic feature has already been noted by \citet{Wyrowski2016} using the NH$_3$ ($J=3_{2+}-2_{2-}$) line, and it is likely related to the expanding shell structure of this source \citep{Peng2010}. The OD $^2\Pi_{3/2}$ ($J = 5/2 - 3/2$) absorption seen towards W49N needs to be fitted with two velocity components; one corresponds to the typically observed narrow absorption feature at the systemic velocity like for 
the other sources, while the other absorption feature is seen over a broad velocity range  of $\Delta \varv$=11\,\kms. This could be explained if it was  associated with outflowing gas at lower velocities than the source's $\varv_{\rm lsr}$.

We detect and resolve two velocity components in absorption towards W51e2, one at the \vlsr\ of the main cloud, and a second component associated with the more diffuse foreground cloud unrelated to W51 Main at $\sim$ 70\,\kms \citep{Mookerjea2014}. This is surprising because this gas has been considered to originate from a layer of diffuse gas in front of the W51 Main complex, unrelated to its star formation activity and should have a much lower column density.

It is clear that OD is predominantly found towards massive envelopes of star forming regions in various evolutionary stages.
However, considering its detection both in the (presumably) more diffuse medium of the 70\,\kms\ component of W51e2 and in outflowing gas from W49N, it appears that it could also be present in sources that have different physical conditions.

\subsubsection{The velocity structure of the line}

Due to the hyperfine structure of the line, the Gaussian fit overestimates the velocity dispersion. Therefore, we also fit the spectra using the GILDAS/CLASS HFS fitting tool that takes into account the hfs splitting and is based on the assumption that the hfs components   {have the same excitation temperature}. This fit constrains the $\varv_{\rm lsr}$ and the line width ($\Delta \varv$) relatively well; however, the line opacity ($\tau$) is poorly constrained due to the blending of the components causing large errors in the opacity estimate. 
To overcome this, we used the line opacity estimated from the line-to-continuum ratio discussed in Sect.\,\ref{sec:abs} and fixed this value for the HFS fit. The values obtained this way are reported in Table\,\ref{tab:results-od}. 

Comparing the results of the HFS fit to those of a simple Gaussian fit, the HFS fitting reveals that the blending of the six hfs components broadens the line widths considerably, typically by up a factor of two compared to a simple Gaussian fit to the blended features.  
In general, the OD absorption is a typically narrow feature with a line width, $\Delta \varv$, between 2.2 and 11.1\,\kms, a mean of 4.2, and a median of 3.2\,\kms. The broadest component is detected towards W49N, and it is likely associated with the outflowing material that is also seen in the \oht\ excited OH line (Sect.\,\ref{sec:exoh}). The emission component seen towards W31C is somewhat broader or blueshifted than the absorption component, suggesting  either a higher level of turbulent broadening compared to the absorption feature  or a different velocity component. This blueshifted component at $\sim-8$\,\kms\ has also been seen in the $o$-NH$_3$ ($J=2_0-1_0$) line at 1214.9\,GHz \citep{Persson2010}. 

The line width from the OD absorption line is, on average, 4.9\,\kms with a median of 3.6\,\kms. This is considerably narrower than that of the C$^{17}$O  ($J=3-2$) line from \citet{Giannetti2014}, which is found to be, on average, 5.7\,\kms (with a median of 5.8\,\kms). The dispersion on the line-width estimates from the C$^{17}$O ($J=3-2$) line exhibit, however, 
a considerably smaller dispersion of 1.7\,\kms, while the OD line widths have a standard deviation of 3\,\kms\ over the sample. This suggests that the OD $^2\Pi_{3/2}$ ($J = 5/2 - 3/2$) line tends to trace a somewhat more quiescent gas compared to that of the C$^{17}$O ($J=3-2$) line.

\subsubsection{OD column density}\label{sec:abs}
The OD $^2\Pi_{3/2}$ ($J = 5/2 - 3/2$) line is detected in absorption towards all sources, and an emission component is only seen towards W31C with a 3.5\,$\sigma_{\rm rms}$ detection measured over the peak brightness temperature of the line. In contrast to the low-mass protostar, IRAS 16293$-$2422 \citep{Parise2012}, the OD absorption towards massive Galactic star forming regions does not appear to be highly saturated; instead, it is typically optically thin. The deepest absorption with $T_{\rm l}/T_{\rm c}$ ratios  of 0.6 and 0.7 is seen towards G23.21$-$0.3 and G351.58$-$0.4, respectively. 

The deepest absorption together with the smallest line-widths are observed towards sources that are typically the youngest, like G23.21$-$0.3, or that are still dominated by a massive, but relatively cold envelope, such as G351.58$-$0.4, G327.29$-$0.6. The total optical depth is $\tau_{\rm tot}=\sum_{i} \tau_{i}$, where $i$ corresponds to the hfs component of the line, and we estimated it using the formula $\tau_{\rm tot}=-\rm{ln} \Big(1-\frac{T_{\rm l}}{T_{\rm c}}\Big)$, where $T_{\rm l}$ corresponds to the main beam brightness temperature of the line and  $T_{\rm c}$ is the continuum temperature accounting for the double-sideband reception of the signal. We find that the values for $\tau_{\rm tot}$ range between 0.04 and 1.29, with a median of 0.15 and a mean of 0.34, implying that the absorption is typically optically thin. The largest optical depths ($\tau_{\rm tot}$) measured towards G351.58$-$0.4 and G23.21$-$0.3 are 1.29 and 0.89, respectively. The smallest values are seen towards the broad absorption component of W49N, and the 70\,\kms\ diffuse cloud component of W51e2.

Adopting the collisional rate coefficients of  OH  also for OD (see Sect.\,\ref{sec:ratran}), we find that the $^2\Pi_{3/2}$ OD $J = 5/2 - 3/2$ line has a very high critical density, $n_{\rm crit}$ of $>10^8$\,cm$^{-3}$. Such a high volume density is unlikely to be  present in the gas in which the OD absorption originates. Therefore, LTE conditions are unlikely to apply. Moreover, the upper level energy ($h\nu/k= 67$\,K) largely exceeds the temperature of the cosmic microwave  background. Therefore, it is safe to assume that  for all sources the vast majority of OD molecules are in the rotational ground state.
The total column density of the molecule can then be estimated as (e.g.\,\citealp{Comito2003}) $N_{\rm tot}=\sum_i \frac{8\pi{\nu_{\rm i}}^{3}}{A_{\rm ul, i} c^3}\Delta\varv_{\rm i}\frac{g_{\rm l, i}}{g_{\rm u, i}}\tau_{\rm i}$, where $\nu_{\rm i}$ is the transition frequency and $A_{\rm ul, i}$ the Einstein-coefficient of transition $i$. $\Delta\varv_i$ is the equivalent width of the line, which relates to its full width at half maximum $\Delta\varv_{\rm FWHM}$ reported in Tab.~\ref{tab:results-oh} via $\Delta\varv=\Delta\varv_{\rm FWHM} \sqrt{\pi/(4\ln{2})}$. $g_{\rm l, i}$ and ${g_{\rm u, i}}$ are the lower and upper state degeneracies, and $\tau_{\rm i}$ is the optical depth at the line centre of the $i$ hfs component. The summation extends over the observed $\Lambda$-doublet component and applies a correction for the OD population in the other one so as to obtain $N_{\rm tot}$. We find values between $3.9\times10^{12}$ and $6.9\times10^{13}$\,cm$^{-2}$, where the lowest values are typically measured towards the more evolved sources  {hosting \uchii\ regions}, and the highest values towards the sources that are dominated by a significant amount of cold dust  {corresponding to quiescent regions and hot cores}. This conclusion is fairly robust against our assumption of a complete ground-state population: we repeated the analysis assuming an excitation temperature of 20~K instead, and the deviation in the deduced $N_{\rm tot}$ is within or close to our fit errors.

\begin{table*}[ht!]
\caption{Fit results of the  $^2\Pi_{1/2}$ OH $J$=3/2--1/2 line. }
\label{tab:results-oh}
\centering
\begin{tabular}{l c c c c c c c c c}
\hline\hline 
   &   $T_{\rm c}^{\rm OH}$$^{(a)}$ & $\sigma_{\rm rms}$ &  $T_{\rm mb}$$^{(b)}$  & $\varv_{\rm lsr}$$^{(b)}$ & $\Delta \varv$$^{(b)}$     \\
   & (K) & (K) &  (K) & (\kms) & (\kms)   \\
\hline
G10.47+0.03   & 5.49 & 0.06          & ~~$1.2\pm0.2 $&  64.3$\pm$0.4 &   4.3$\pm$1.1 \\
                &        &                         & $-0.8\pm0.1$  &  49.5$\pm$1.7 &   26.9$\pm$3.0 \\
G34.26+0.15$\dagger$  &  8.66   & 0.14   & $1.6\pm0.2$ & 58.1$\pm$0.3   &   4.2$\pm$0.7   \\
                    &                &                   &  ~~$-1.1\pm0.3$   &  62.5$\pm$0.3   &   1.5$\pm$0.6   \\
G35.20$-$0.7  &  2.02 & 0.15    & $2.1\pm0.2$  &  34.5$\pm$0.4 &   9.7$\pm$0.9       \\
W28A      & 6.80   & 0.11 & $5.7\pm2.1$ & 26.0$\pm$6.2 & 34.2$\pm$ 8.5  \\
                &               &                        & $4.0\pm1.5$ & 12.4$\pm$1.3  &  13.7$\pm$5.8 \\
W31C   &   8.79 & 0.12           & $3.5\pm0.3$  &  $-$3.3$\pm$0.5 &  7.8$\pm$1.0 \\
             &           &                        &  $2.0\pm0.3$  &   4.2$\pm$1.2 &   14.9$\pm$2.9   \\
W33A    &  2.32  & 0.13   & $1.5\pm0.1$ & $40.7\pm0.4$ & $16.3\pm1.2$\\
W49N   & 12.35   & 0.07    & $4.8\pm0.4$ & $19.8\pm0.4$ & $10.2\pm1.1$ \\
       &         &            & $3.8\pm0.3$ & $6.7\pm1.2$ & $12.8\pm1.2$ \\
       &         &            & $2.9\pm0.2$ & $13.9\pm3.8$ & $68.6\pm3.8$ \\
G327.29$-$0.6 &  5.53 & 0.06   & -- & -- & -- &\\
G351.58$-$0.4 & 6.17 & 0.10   &  0.7$\pm$0.2 & $-86.0\pm1.9$ &  15.9$\pm$3.9& \\
\hline
\end{tabular}
\tablefoot{{$^{(a)}$} Main-beam brightness temperature of the single-sideband continuum at 1.838\,THz (Rayleigh-Jeans scale). The velocity resolution is 0.75\,\kms. {$^{(b)}$} The errors are constrained by the fitting procedure in CLASS.
$\dagger$The hfs fit is for the 1834.8\,GHz component of the $\Lambda$-doublet from the image band. \\}
\end{table*}

\subsection{The  $^2\Pi_{1/2}$  OH ($J$=3/2--1/2) transition}\label{sec:exoh}

We observed the rotationally excited  $^2\Pi_{1/2}$ OH ($J$=3/2$-$1/2) transition at 1834.7\,GHz towards ten sources of the sample, and show both $\Lambda$-doublet transitions in Fig.\,\ref{fig:oh_all}. Two sources without an OD detection, W51d and G31.41+0.2, as well as the quiescent source, G23.21$-$0.3, and the {\uchii} region, W51e2, have not been observed in this line. Furthermore, the excited OH line is not detected towards G351.58$-$0.4 and G327.29$-$0.6. Towards W33A, only one of the $\Lambda$ doublets falls in the band.

The remaining sources show relatively bright emission in the $^2\Pi_{1/2}$ OH ($J$=3/2$-$1/2) line. 
The $\Lambda$ doublets have three hfs components  {(Table\,\ref{tab:lines})}, the 1834.7\,GHz line has an isolated hfs component separated by 11.5\,\kms, which appears as a distinct feature in the spectra. The largest velocity shift for the other  $\Lambda$-doublet component at 1837.8\,GHz is separated by 1.9\,\kms, so all its three hfs components appear blended in the spectra.

The OH line profiles frequently exhibit a broad velocity component likely associated with outflowing high-velocity gas, which is very pronounced in W49N, W28A, and W33A. Towards G10.47+0.03, the outflow component appears to be in absorption. Previously, outflowing gas has been found in excited OH towards the high-mass star forming region W3 IRS 5, where \citet{Wampfler2011} also detected both the envelope and outflow components. Our previous SOFIA/GREAT observations of this excited OH line did not have a high enough signal-to-noise ratio to be sensitive to weaker broad components \citep{Csengeri2012}. Low spectral resolution observations with \textit{Herschel} similarly suggest that the exited OH line can be frequently associated with emission from the high velocity outflowing gas (cf.\,\citealp{YLYang2018}). The high-velocity outflowing gas of W28A has been studied in detail, including the OH data  shown here, in \citet{Leurini2015}. 

In addition to the broad velocity component, a narrow component associated with the envelope is always present. Furthermore, the line profiles towards G34.26+0.15 suggest that both the emission and absorption components are present in the excited OH line, similarly to what is seen in other high density gas tracers (cf.\,\citealp{Wyrowski2016}). The spectra towards G10.47+0.03 are a mixture of a narrow emission component and a broader absorption component from the outflowing gas, as discussed above. The double-peaked profile towards W49N suggests  {absorption from lower-excitation gas in the envelope}. 

Similarly to our previous work in  \citet{Csengeri2012}, we used the HFS method of CLASS to fit the lines.  {We did this separately for the $\Lambda$ doublets in the signal and image side band, respectively, and list the average of these results  
in Table\,\ref{tab:results-oh} and display them in Fig.\,\ref{fig:oh_all}}.  {Except for W49N, where three components were needed, typically one narrow and a broader component was sufficient to fit the data. The brightest emission is seen towards sources that appear typically weak in the ground-state OD line.}

The average FWHM line width for the $^2\Pi_{1/2}$ OH ($J$=3/2$-$1/2) line is 10.3\,\kms\ for all the sample considering the narrowest components, where two velocity components were fitted. This is more than two times larger than that of the OD $^2\Pi_{3/2}$ ($J = 5/2 - 3/2$) line, where for the sources also observed in the $^2\Pi_{1/2}$ OH ($J=3/2-1/2$) line we find an average line-width of  3.6\,\kms.

\section{Line modelling with RATRAN}\label{sec:ratran}

\subsection{The physical structure of the envelope}
To estimate the abundance distribution of OD and to compare it to that of OH throughout the envelope, we use the non-LTE radiative transfer code RATRAN \citep{HvdT2000}. Following the work of \citet{Wyrowski2016}, we 
constrained the physical structure of the envelope  (such as size and the density profile) with the 870\,$\mu$m dust emission as observed in the APEX Telescope Large Area Survey of the Galaxy \citep[ATLASGAL,][]{Schuller2009}. This implies that we assume that the densities are high enough for the gas and dust to be  {in thermal equilibrium}, which is a reasonable assumption for high-mass protostellar envelopes embedded in massive clumps with high densities. The corresponding parameters of the physical properties are listed in Table\,\ref{tab:results} for sources with a detection in the ground-state OD line. This  allowed us to model six sources. 

Our models use a spherical geometry with a power-law profile for the density, $n$, as a function of radius, $r$, given by  $n(r)\sim r^{\alpha_n}$, and a temperature, $T$, profile of $T(r)\sim r^{-0.4}$. The line width  
is obtained from the C$^{17}$O (3--2) data for  these sources \citep{Giannetti2014}. The velocity gradient through the envelope is constrained by NH$_3$ observations and the models of \citet{Wyrowski2016}. The physical structure of the model for G34.26+0.15  is illustrated in Fig.\,\ref{fig:model},  where the bottom panel shows the modelled abundance profiles for OH and OD (see Sect.\,\ref{sec:ratran_g34}). We used the recent collisional rate coefficients from \citet{Klos2017} for OH for collisions with H$_2$, and adopted these coefficients for OD (see App.\,\ref{app:od_coeff} for the validity of this approach).
We then varied the OH and OD abundances ($X$(OH) and $X$(OD)), respectively.

Radiative heating from (high-mass) protostars leads to a radial increase in temperature towards the centre of the envelope \citep[e.g.][]{WolfireCassinelli1986,GoldreichKwan1974}. At low temperatures ($< \sim 20$~K), the chemistry is determined by grain surface processes. Triggered by the ignited young stellar object, molecules are desorbed. This greatly changes gas-phase abundances, which are further modified by gas-phase reactions. More importantly, deuterium fractionation of several species is expected to be different in gas-phase reactions compared to grain surface processes (e.g.\,\citealp{Roberts2007}), leading to different molecular abundances in the inner (hot) and outer (cold) regions of envelopes. Since the grain mantles are expected to be mostly composed of water ice, the typical temperature (and thus radius) determining the change of abundances is related to the water evaporation temperature around 100\,K \citep{Fraser2001}.

Using 1D, non-LTE radiative transfer modelling, our aim is to  estimate the OD abundance in the envelope and to probe what fraction of the OD originates from the hot inner regions and the cold outer envelope. Then, assuming that OH and OD are part of the same gas, we constrained their abundances throughout the envelope and thereby determined the deuterium fractionation by $D^{OD}=X(\rm{OD})/X(\rm{OH})$.

For our models, the ground-state \oht\ OH line seen in absorption should, in principle, provide the best constraint on the OH abundance in the outer envelope. However, as discussed in detail by \citet{Wiesemeyer2016}, this line is largely optically thick and is severely blended with absorption from diffuse gas along the line of sight (cf.\,\citealp{Wiesemeyer2012, Wiesemeyer2016}). Therefore, we can only check whether our models are consistent with these observations. However, towards G34.26+0.15 
we obtained archival data from SOFIA covering the  \oht\ $^{18}$OH ($J=5/2-3/2$) ground-state transition at 2494.6950\,GHz.  This line shows an emission spike at $\sim$50\,\kms\  due to a telluric line, which, against strong continuum sources, makes the calibration locally challenging (see Figs.\,\ref{fig:g34} and \citealp{Wiesemeyer2012} for details). Nevertheless, we also included the $^{18}$OH  line in our models assuming an isotopic ratio of $^{16}$O/$^{18}$O=400 \citep{Wilson1999}. 

The \oho\ OH ($J=3/2-1/2$) excited-state transition has an upper energy level of 270\,K compared to the 67\,K of the \oht\ OD ($J=5/2-3/2$) transition, and so it is clearly a more sensitive probe of the warmer inner regions. Nevertheless, we also consider this line to further constrain our models, allowing us to  estimate the total OH column density in the envelope.

\subsection{G34.26+0.15: Non-LTE modelling of OH, OD, HDO, and H$_2^{18}$O using RATRAN}\label{sec:ratran_g34}
\subsubsection{Modelling constraints}
 The source G34.26+0.15 exhibits a profile  with redshifted self-absorption in the excited-state OH line, and its source structure is relatively well constrained.  Hence, it is for this source that we can perform a more detailed modelling  in order to test different OD abundance profiles. The simplest model ($A$) corresponds to a constant abundance profile through the envelope, while models $B, C$,and $D$ introduce an abundance jump of OD above a  temperature ($T_{\rm J}$) of 50, 100, and 200\,K,  respectively (Fig.\,\ref{fig:model}, bottom panel).  OD originating from the warm inner envelope is expected to show up in emission in the blueshifted part of the spectrum, allowing us to constrain the deuterium fractionation in  this source for both the inner, warm, and outer cold envelope (appearing in absorption). We list the model parameters, radius ($r_{\rm J}$), and temperature ($T_{\rm J}$) corresponding to the abundance jump, together with the range of abundances and the modelling results, in Table\,\ref{tab:g34-model}.

 For model \textsl{A}, we created a grid of parameters for the OH abundance, $X(\rm{OH})$, between $10^{-10}$ and $1\times10^{-8}$.  We modelled the deuterium fractionation $D^{OD}=X(\rm{OD})/X(\rm{OH})$ between $10^{-4}$ and $5\times10^{-1}$. These ranges were sampled by 20 points on a logarithmic scale.  For models \textsl{B} and \textsl{C,} we used the result from model \textsl{A} to fix the OH abundance and the deuterium fractionation in the outer envelope, and we varied $D^{\rm{OD}_{\rm in}}$ as listed in Table\,\ref{tab:g34-model} in an inner radius between $r_{\rm J}$=0\rlap{.}{\arcsec}2 and 5\arcsec,\ which translates to physical scales of $\sim$320 and 8\,000\,au, respectively.
After performing the calculations with RATRAN, we convolved the modelled spectra with the beam of the telescope at the respective frequencies shown in Table\,\ref{tab:lines} and the hfs structure assuming LTE conditions, meaning we adopted the relative hfs intensities according to their rates of spontaneous emission or absorption\footnote{This approximation is not valid for the highly optically thick absorption seen in the ground-state \oht\  {OH} line.}. The models have been compared to the observations and were evaluated by a reduced $\chi^2$ method.

For the other molecules shown  {in Fig.\,\ref{fig:g34}, we used the same physical model as for OH and OD, and we used the abundance of \citet{Wyrowski2016} for NH$_3$}, and   a constant abundance profile for HDO and H$_2^{18}$O. Since we used the revised distance estimate of 1.4\,kpc based on maser parallax measurements \citep{Kurayama2011} for G34.26+0.15,  compared to \citet{Coutens2014}, we  find that we need to scale  their abundances  of HDO and H$_2^{18}$O for the outer envelope by factors of 5 and 1.5, respectively, in order to obtain a good fit to the data\footnote{This gives a deuterium fractionation for water vapour in the outer envelope of $5.1\times10^{-3}$ instead of $1.5\times10^{-3}$ reported in \citet{Coutens2014}. }. 

\begin{figure}[!t]
   \centering
   \includegraphics[width=0.9\linewidth]{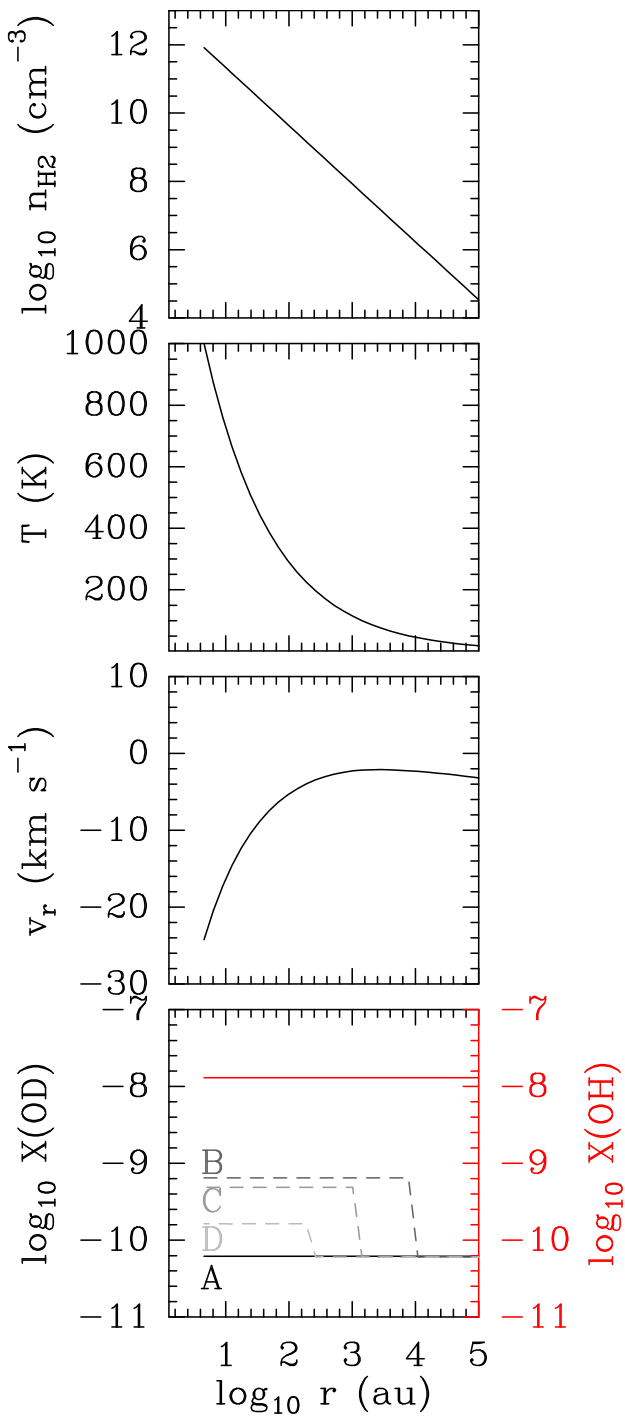}
      \caption{Physical structure of models for G34.26+0.15. Bottom panel shows the OH (\textit{red}) and OD (\textit{grey}) abundance variations in the envelope  {for the different models}.  For a better visibility, the OD abundance in the inner envelope ($X$(OD)$^{\rm in}$) is shifted for the different models. All models test the same abundance range.}
              \label{fig:model}%
    \end{figure}
\begin{figure}[h]
   \centering
      \includegraphics[width=0.95\linewidth]{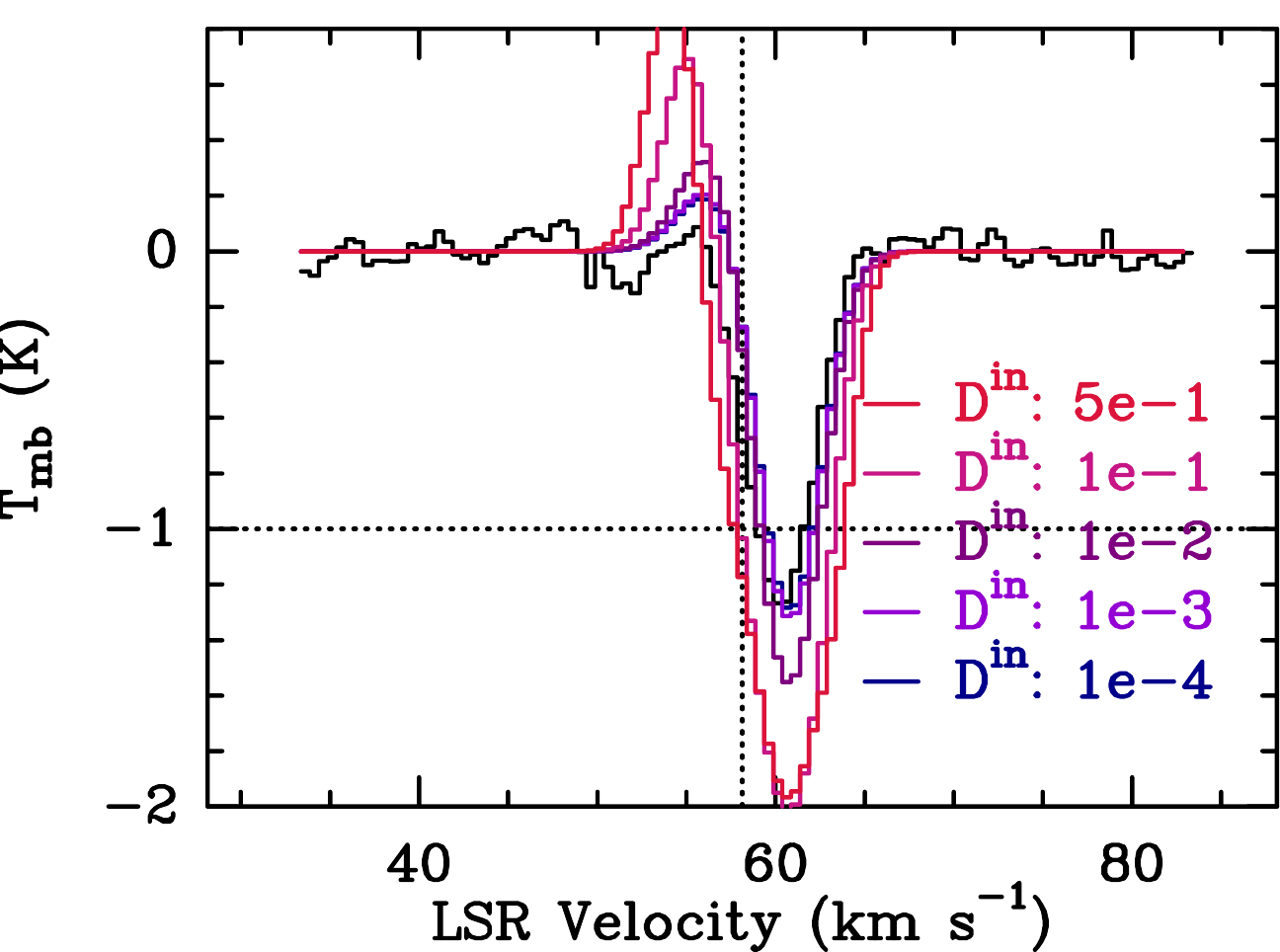}

      \caption{Results of model \textsl{B} for  $^2\Pi_{3/2}$ OD ($J$=5/2--3/2) transition of G34.26+0.15 using $T_{\rm J}=50$\,K. The different $D_{\rm in}^{\rm OD}$ values used for the models are shown in the figure legend.}
              \label{fig:B}%
    \end{figure}
\begin{figure*}[!htpb]
   \centering
   \includegraphics[width=\linewidth]{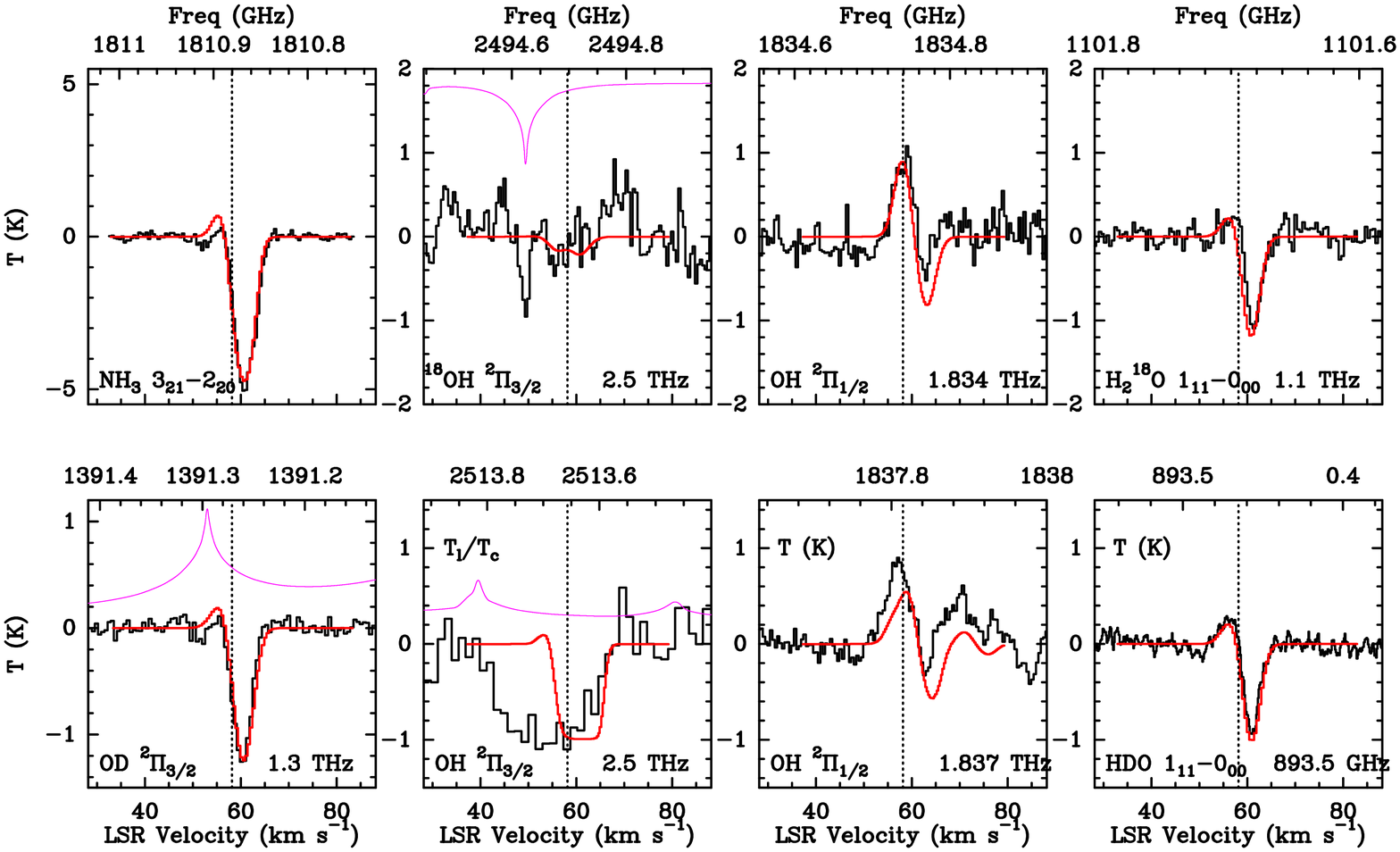}
   \caption{Observed spectra towards G34.25+0.15 are shown in black, and the red line shows the fitted RATRAN model ( {model {\sl A}}). Dotted line marks the $\varv_{\rm lsr}$ of the source. Magenta line shows the atmospheric transmission, where  
   {the absorption feature in the $^{18}$OH spectrum around 50\,\kms\ is due to a deliberately uncorrected telluric ozone feature.}
   }
              \label{fig:g34}%
    \end{figure*}

\subsubsection{Constant {deuteration profile towards G34.26+0.15}}

Our first result  for G34.26+0.15 is that already a constant OH and OD abundance profile throughout the entire envelope (model $A$) gives a good fit to all the lines shown in Fig.\,\ref{fig:g34}. Our best fit  gives an OD abundance, $X$(OD), of $6.2\times10^{-11}$, and we find that an OH abundance, $X$(OH), of $1.3\times10^{-8}$ is consistent with the observed \oht\ OH ground-state absorption and the \oho\ excited-state line\footnote{
Our model over-predicts the 1.3\,THz continuum for the OD $^2\Pi_{3/2}$ $J$=3/2--1/2 transition by 0.65\,K (corresponding to 10\%), and it under-predicts the continuum by 0.9\,K (10\%) at 1.8\,THz for the OH $^2\Pi_{1/2}$ $J$=5/2--3/2 line. It  shows a discrepancy of 
3.0\,K (corresponding to 50\%) for the OH $^2\Pi_{3/2}$ $J$=3/2--1/2 line at 2.5\,THz.
}. This   model corresponds to a constant deuterium fractionation of $4.7\times10^{-3}$  throughout the envelope. Our fit reproduces the 1834\,GHz $\Lambda$-doublet pair of the \oho\ OH excited line relatively
well, although it somewhat over estimates the absorption component. On the other hand, the fit to the 1837\,GHz $\Lambda$ doublet is less good: it underestimates the emission component, and, as for the ground-state-line $\Lambda$-doublet pair, it overestimates the absorption component.

Using the same physical model, we also fitted the HDO and p-H$_2^{18}$O ($J=1_{1,1}-0_{0,0}$) lines with a constant abundance profile of $X$(HDO)=$1.6\times10^{-11}$ and $X$(H$_2^{18}$O)=$7.8\times10^{-12}$ (corresponding to $X$(H$_2$O)=$3.1\times10^{-9}$ assuming an isotopic ratio of $^{16}$O/$^{18}$O=400). These lines were modelled in detail by \citet{Coutens2014}, who adopted a distance of 3.3\,kpc for the source, almost two times larger than what we used here based on \citet{Wyrowski2016}. It is already interesting that we obtain a good fit to these lines without an abundance jump, suggesting that these observations are not sensitive to the hot inner regions. However, our models require an $X$(H$_2$O) more than three times lower compared to \citet{Coutens2014} for the outer envelope. In Fig.\,\ref{fig:g34}, we also show the observations and the model of the NH$_3$ $J=3_{2,1}-2_{2,0}$ line from \citet{Wyrowski2016} as a comparison. Our results therefore suggest a deuterium fractionation of $D^{\rm OD}_{\rm out}=4.7\times10^{-3}$ for OD, very similar to that of HDO/H$_2$O of $D^{\rm {HDO}}_{\rm out}=5.1\times10^{-3}$  {constrained by our modelling here}.

\subsection{Impact of evaporation}
Due to  water ice evaporation, a change in the abundance profile and deuterium fractionation can be expected (see also \citealp{Coutens2014} and Sect.\,\ref{sec:photo}). Therefore, we also tested models introducing a change in deuterium fractionation at a given temperature ($T_{\rm J}$). In models $B$, $C$, and $D$ we varied the temperature of the abundance jump ($T_{\rm J}$=50,100, 200\,K) and varied the deuterium fractionation in the hot inner regions compared to the outer envelope ($D_{\rm in}^{\rm OD}$) in a range of $10^{-4}-5\times10^{-1}$. 

In Fig.\,\ref{fig:B}, we show the results of our model $B$ with $T_{\rm J}=50$\,K that introduces the abundance jump at a few thousands of au in the envelope, which is comparable to the physical scale of water evaporation found by \citet{Coutens2014}. 
 These models show a noticeable change in the blueshifted component of OD seen in emission as a result of the velocity gradient over this source. 
As we see below, given our sensitivity limit, this is the only model that can put a constraint on the OD column density in the inner regions at $r_{\rm in}\sim11\,000$\,au. Our results suggest that the observations  are consistent with a  deuterium fractionation in the inner  envelope ($D^{\rm OD}_{\rm in}$) of up to $\sim5\times10^{-3}$ , which would imply that a jump of deuterium fractionation up to a factor of 2 could remain undetected in our observations.  However, a strong deuterium enhancement towards the inner regions is clearly excluded by these models.
This suggests that the OD abundance in the inner regions cannot be significantly higher than in the outer regions, and thus it does not favour  an enhancement in deuterium fractionation in the inner regions compared to the cold envelope. 
A firmer constraint on the deuterium fractionation towards the innermost regions can only be obtained by observations of higher energy transitions of OD. 

We notice that models $C$ and $D$ do not introduce any noticeable change in the model results  compared to model \textsl{A}, {and additional calculations show that for  `jump' temperatures $T_{\rm J}{\geq}60$\,K, the hot emitting region becomes too small in angular size (${\leq}$3\arcsec corresponding to 4800\,au) and their emission or absorption is heavily diluted in the SOFIA beam. 

 We also find that at abundances slightly higher than the values we investigate here, the emission of OH and OD becomes optically thick, and hence it cannot provide a view of the inner regions. Instead, as the higher energy transitions get excited, they should provide a better probe of the inner regions. This fact is somewhat compensated by the velocity structure of the source for G34.26+0.15 and G351.58$-$0.4 (Fig.\,\ref{fig:model}), allowing us to probe deeper into the envelope due to the accelerated infall, which results in  red-shifted velocities closer to the source. For G34.26+0.15 (in contrast to \citet{Coutens2014}, whose best fitting models use a temperature of 200\,K for an abundance jump), due to the  {revised distance for this source}, our models at the same temperature correspond to a considerably more compact hotter region within 270\,au. The \citet{Coutens2014} best fit corresponds to an extent of 1.7\arcsec (thus 6600\,au) in their models, meaning to a value similar to that of our low $T_{\rm J}$ models.

\begin{table*}[ht!]
\setlength{\tabcolsep}{3pt}
\caption{Fit parameters from RATRAN modelling of the OH and OD lines.}
\label{tab:results}      
\begin{center}          
\begin{tabular}{l c c c c c c c c c c c}
\hline\hline       
Source  & $d$&  $L_{\rm bol}$ & $R_{\rm out}$   &       $\alpha_{ \rm n}$       & $n_{\rm 1\,pc}$ & $\Delta\varv$ &       $X$(OH) & $X$(OD)       & $D^{\rm OD}_{\rm out}$ &  $D^{\rm HDO}_{\rm out}$ & Ref. \\
                & (kpc) & ($L_{\odot}$) &       (pc)                    &                               & ($10^{3}$ cm$^{-3}$)    &       (km\,s$^{-1}$)  &                       & & & & for       $D^{\rm HDO}$   \\
\hline
G23.21-0.3  & 4.6 & $1.3\times10^{4}$   & 1.8           &       $-2.0$          &  4.5                    &       1.0             &         --  & $8.0\times10^{-10}$     & --      &  -- & -- \\
G34.26+0.2 & 1.6 & $4.6\times10^{4}$ &  0.8                     &       $-1.8$          &  10                     &       2.4             &       $1.3\times10{^{-8}}$    &       $6.2\times10^{-11}$     & $4.8\times10^{-3}$       & $5.1\times10^{-3}$ & (a)\\
& & & & & & & & & & $1.0-3.1\times10^{-3}$ & (b,c)\\\
&   &                           &                       &               &                         &       &        &   & & $1.9-4.9\times10^{-4}$ & (d)\\
G34.41$+$0.2 & 5.4 & $6.7\times10^{4}$ & 2.5 & -2.0 & 7 & 3.2 & -- & $<8.0\times10^{-11}$ & -- & -- & -- \\
G35.20$-$0.7 & 2.2 & $2.5\times10^{4}$ & 1.5 & $-1.6$ & 5.5 & 1.5 & $>2.0\times10^{-8}$ & $3.0\times10^{-10}$ & $<1.5\times10^{-2}$ & -- & -- \\
G327.29$-$0.6 & 3.1     &  $8.2\times10^{4}$    & 2.0           &       $-1.9$          &  10                     &       2.3             &               $<5\times10^{-9}$       &       $9.75\times10^{-11}$    & $>2.0\times10^{-2}$ &$5.9-6.4\times10^{-3}$ & (e) \\
G351.58-0.4 & 6.8 & $2.4\times10^{5}$   & 1.8           &       $-1.9$          &  15                     &       1.5     & $7.8\times10^{-9}$    &$1.95\times10^{-10}$ & $2.5\times10^{-2}$    &       $0.6-1.4\times10^{-3}$ & (e)                    \\
\hline

\hline
\end{tabular}
\end{center}          
\tablefoot{Outer radius, $R_{\rm out}$, density power-law index, $\alpha_{\rm n}$, and the density at 1\,pc, $n_{\rm 1\,pc}$, were constrained by the 870\,$\mu$m ATLASGAL dust continuum emission and are adopted from \citet{Wyrowski2016}, as are the distances.\\ 
$^{a}$ \citet{vanderTak2006}. 
(a) This work. (b) \citet{Coutens2014}. (c) \citet{Jastrzbska2016}. (d) \citealt{Liu2013}. (e) \citet[PhD thesis][]{Liu_thesis}.
}

\end{table*}

\begin{table*}[ht]
\setlength{\tabcolsep}{3pt}
\caption{Summary of RATRAN model parameters and results towards G34.26+0.15.}
\label{tab:g34-model}      
\begin{center}
\begin{small}
\begin{tabular}{l c c c c c c c c r l}
\hline\hline 
Mod.    & {$X$(OD) } & $r_{\rm J}$ & $X$(OH) &$X$(OD)${^\dagger}$ & $D_{\rm in}^{{\rm OD}}$ & $D_{\rm out}^{\rm {OD}}$\\

                & profile & (\arcsec)&  model range (result)&   model range (result)               &    \\
\hline
A  & {constant}  & -- & $10^{-10}$--$1\times10^{-8}$ (${\bf{1.3\times10^{-8}}}$) & $1.0\times10^{-14}$--$2.5\times10^{-9}$ (${\bf{6.2\times10^{-11}}}$) & -- & ${10}^{-4}$--$5\times10^{-1}$ (${\bf{4.8\times10^{-3}}}$) \\
B &  $T_{\rm J}=50$~K &  ~~5.0\arcsec & $1.3\times10^{-8}$ & $1.3\times10^{-12}-6.5\times10^{-10}$ & $10^{-4}-5\times10^{-1}$ (${\bf{\leq5\times10^{-3}}}$) & ${{4.8\times10^{-3}}}$\\
C &  $T_{\rm J}=100$~K & 1\rlap{.}{\arcsec}0 & $1.3\times10^{-8}$ & $1.3\times10^{-12}-6.5\times10^{-10}$  & $10^{-3}-5\times10^{-1}$${^\ddagger}$  & ${{4.8\times10^{-3}}}$\\ 
D &  $T_{\rm J}=200$~K & 0\rlap{.}{\arcsec}2 &
$1.3\times10^{-8}$ & $1.3\times10^{-12}-6.5\times10^{-10}$ & $10^{-3}-5\times10^{-1}$${^\ddagger}$  & ${{4.8\times10^{-3}}}$\\ 
\hline
\end{tabular}
\end{small}
\end{center}          
\tablefoot{Bold face values correspond to the best fit results. ${^\dagger}$ $X$(OD) corresponds to $X$(OD)$^{\rm in}$ for models \textsl{B}, \textsl{C}, and \textsl{D}. ${^\ddagger}$ Due to the small source size, all these models are consistent with the data.}
\end{table*}

\subsection{Other sources}

Above, we discuss the radiative transfer modelling of G34.26+0.15 in detail. For the other sources, we refrained from such a comprehensive modelling in this work due to the lack of a similarly detailed knowledge about the source structure and the lack of sufficient ancillary data in the form of HDO and H$_2^{18}$O lines. For some sources, though, we were able to rely on the  source model of \citet{Wyrowski2016}, and we used this as an input for RATRAN to obtain column density estimates primarily for OD and, where available, for OH based on the excited $\Lambda$ doublets.  {All these models assume a constant OD abundance profile throughout the envelope  {similarly to model \textsl{A} for G34.26-0.15}.}

The parameters of \citet{Wyrowski2016} use the C$^{17}$O (3--2) measurements from \citet{Giannetti2014} to define the sources' LSR velocity. They provide a good model for  NH$_3$. We adopted the same line width for the OD ground-state absorption, which worked well for all sources, except for G351.58$-$0.4, for which  we find a two times larger line width from the HFS fitting. Our models require this broader line component to provide a good fit to the  OD spectrum  {(see also App.\,\ref{app:g351})}. Similarly, G35.20--0.7 also requires a larger line width (of 2.6\,\kms)\ than the previous models; however, in this case we find that a constant OH abundance profile does not give a good fit to the excited OH line. The line profiles suggest self-absorption, and while our models reproduce the line profile, they typically underestimate the line intensity,  {suggesting that OH could be present in a higher abundance towards the warmer inner regions}. We suggest that an increased OH column density in the inner envelope could provide a better fit to the observations. The reported OH column density is therefore an upper limit for the outer envelope, and by increasing its value even the excited OH line would appear only in absorption.
 Due to the limitations of the models for the OH abundance, we obtain the best constraints on the deuterium fractionation ($D^{\rm OD}$) for G34.26+0.15.}The results of these models are summarised in Table\,\ref{tab:results}. 

Altogether, the non-LTE radiative transfer modelling gives constraints on $X(\rm{OD})$ towards six sources of the sample. These results agree within a factor of 3 with the 
ones calculated from the absorption features in Sect.\,\ref{sec:od_ground_state}. The largest differences are seen towards G23.21-0.3 and G327.29$-$0.6, where the modelling gives nearly three times higher OD abundances. 

   \begin{figure}[!htpb]
   \centering
   \includegraphics[width=0.9\linewidth]{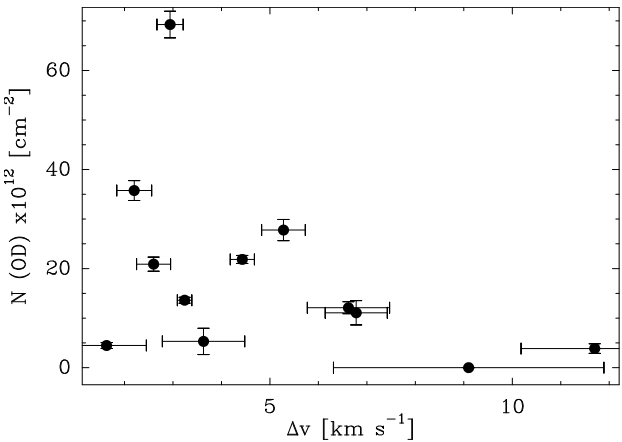}
      \caption{Total column density ($N$) of the ground-state $^2\Pi_{3/2}$ OD ($J = 5/2 - 3/2$) line versus the line width ($\Delta v$) obtained from the HFS fitting with a fixed total optical depth from the line-to-continuum ratio.}
              \label{fig:dv_ncol}%
    \end{figure}

\section{Discussion}\label{sec:discussion}
\subsection{OD absorption from the cold outer envelope}


Absorption from the \oht\ OD ($J = 5/2 - 3/2$) ground-state line seems to be prevalent in high-mass star forming regions. It is, however, not exclusively associated with the high column density envelopes, but it is also observed towards various physical components within these regions. It is present in the broad absorption from the outflowing high-velocity gas towards W49N, as well  as  in the more diffuse gas component seen, for example, towards the 70\,\kms\ cloud component around W51e2. In contrast to the low-mass protostar IRAS 16293$-$2422 \citep{Parise2012}, where the absorption is close to saturation, here the largest optical depths are seen towards the youngest, quiescent sources, and nearby hot cores, where a significant amount of cold dust is still present. These findings are consistent with the chemical models for OD by \citet{Croswell1985} that predict a high fractionation of OD in the low-density cold gas, while in the dense gas even higher fractionation could be expected depending on the local conditions, such as UV field and fractional ionisation.

Besides the potential outflow and diffuse gas origin, we focus here on the origin of OD within the high-mass protostellar envelopes, that is, the dense gas component. After accounting for the hfs splitting, the line widths we determine  are narrow, which is consistent with an origin in the outer envelope where the gas is quiescent. Fig.\,\ref{fig:dv_ncol} shows a comparison between the measured line-width ($\Delta v$) from the OD absorption compared to the  estimated total column density. The largest line widths are associated with a more turbulent component of the gas that seems to be less optically thick and have the smallest OD column density in the sample. The largest OD columns show narrow line widths, suggesting an origin from the dense gas of the envelope.

Comparing the $\varv_{\rm lsr}$ and $\Delta \varv$ of the \oht\ OD ground-state line with  those of the NH$_3$ ($J=3_{2+}-2_{2-}$) absorption at 1.8\,THz from \citet{Wyrowski2012, Wyrowski2016}, we notice that the line widths and the line shapes are in general similar (see App.\,\ref{app:nh3} for more details). This further suggests a common origin in the cold absorbing layer of the envelope. Comparing with NH$_3$, in two cases we find a considerably larger line width in the OD line than in the NH$_3$ lines, suggesting a potential contribution from the more inner layers of the envelope. However, there are also other noticeable differences compared to the 1.8\,THz NH$_3$ line. In contrast to the narrow OD absorption, NH$_3$ may exhibit velocity components that are not seen in the $^2\Pi_{3/2}$ OD ($J = 5/2 - 3/2$) line. For example, the broad absorption components seen in NH$_3$ towards the sources G327.29$-$0.6 and G351.58$-$0.4 are less pronounced in OD, suggesting either a chemically different environment, or different excitation conditions for this gas component.  
 
The strongest confirmation that the $^2\Pi_{3/2}$ OD ($J = 5/2 - 3/2$) line originates from the cooler layers of the envelope is given by our detailed radiative transfer modelling of G34.26+0.15. We find that the outer layers of the envelope at typically low temperatures ($T_{\rm d} < 80-100$\,K) and densities below $\sim10^{7}$\,cm$^{-3}$ are responsible for the absorption feature (Fig.\,\ref{fig:model}). The innermost regions are affected by optical depth effects, the modelled transitions of both OH and OD become optically thick, even before the gas density reaches the critical density ($n_{\rm cr}$).  

With respect to the elemental D/H ratio ($\sim1.5\times10^{-5}$, \citealp[e.g.][]{Vidal-Madjar1983, Linsky2003}), chemical models predict a considerable enhancement of OD in the diffuse gas \citep{Croswell1985}, where most of the deuterium is expected to be present in atomic form. A significant amount of OD can then be formed through rapid exchange reactions \citep{Roberts2002b}; the reaction D + OH {$\rightleftharpoons$} H + OD is exothermic and has a reaction rate coefficient that is higher at low temperatures \citep{Margitan1975,Atahan2005, Wang2006}, leading to a stronger enhancement in OD towards colder gas. 
The destruction of OD mainly happens through reactions with C$^+$ ions and photodissociation. Our values for the deuterium fractionation in OD, ranging between 4.8$\times$10$^{-3}$ and 2.5$\times$10$^{-2}$, are higher than those predicted for the diffuse gas by these chemical models, and are consistent with a picture where in the dense gas, with less C$^+$ and UV photons the destruction  of OD slows down, leading to a stronger deuterium fractionation compared to the diffuse gas. 

\begin{figure}[!htpb]
   \centering
   \includegraphics[width=\hsize]{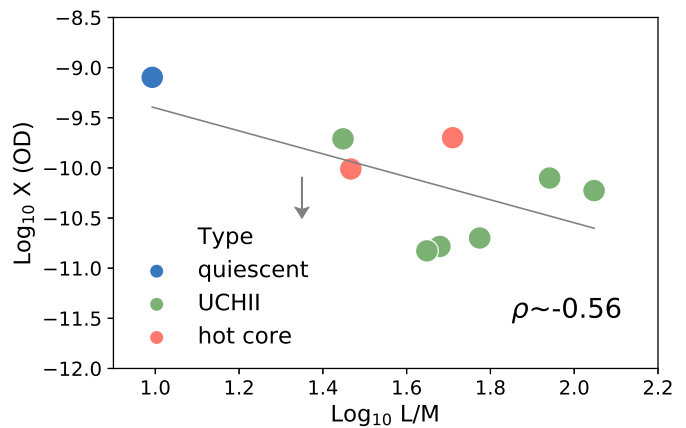}
    \caption{Observed OD abundance ($X$(OD)) versus $L_{\rm bol}/M$ ratio for the entire sample. The colors show the different source types. The grey arrow marks the upper limit estimated for the non-detection of the quiescent source, G34.41. The Pearson correlation coefficient ($\rho$) is labelled in the figure.}
    \label{fig:XOD_LM}
     \end{figure}
\begin{figure}[!htpb]
   \centering
   \includegraphics[width=\hsize]{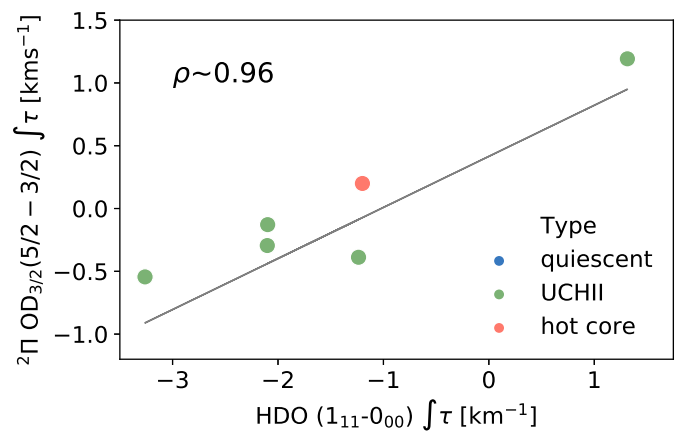}
    \caption{Observed OD and HDO line area measured on the line-to-continuum ratio. The colours show the different source types. }
    \label{fig:od-hdo-corr}
     \end{figure}

\subsection{Decreasing OD abundance towards the more evolved sources}

To investigate the OD abundance variation over our sample of massive clumps in various evolutionary stages, we first used our values  for the OD column density, $N_{\rm OD}$, from Table\,\ref{tab:results-od} towards all sources with a detection in the \oht\ OD ground-state line. We estimated the H$_2$ column density from the 870\,$\mu$m beam averaged flux density from the ATLASGAL point source catalogue \citep{Csengeri2014}, because the APEX beam at this frequency is comparable to the beam size of 20\arcsec,\ corresponding to the OD observations. For the dust temperature ($T_{\rm dust}$), we used the values estimated by \citet{Konig2017} based on fitting the spectral energy distribution (SED) of the dust emission. The $T_{\rm dust}$ values are found to be between 22.1 and 34.5~K, the temperatures around 30\,K are found towards sources hosting embedded {\uchii} regions. Then, we computed the observationally estimated OD abundance by $X(\rm{OD})$=$N_{\rm OD}$/$N_{\rm H_2}$. 

We complemented the observationally derived $X(\rm{OD})$ estimates (where available) with the results of the radiative transfer modelling (Sect.\,\ref{sec:ratran}), and hence we can discuss the abundance variations for a sample of  nine sources. We find that the highest OD abundances are near the youngest sources of the sample, except for the non-detection towards G34.41+0.2. In particular, G23.21-0.3 and G35.20$-$0.7  show about an order of magnitude larger abundances than the rest of the sample.  Fig.\,\ref{fig:XOD_LM} shows the distribution of $X(\rm{OD})$ as a function of $L_{\rm bol}/M$ that we use as a proxy for the evolutionary stage of the clump (see, however, the work of \citealt{Ma2013}  {about $L_{\rm bol}/M$ reflecting the mass of the central object}). Pearson's correlation coefficient ($\rho$) is $-0.56$, suggesting a trend of decreasing $X(\rm{OD})$ over $L_{\rm bol}/M$. Due to the limited number of sources observed, this correlation is, however, not statistically significant (false alarm probability 16\%).

We find that the classification of the sources based on the diagnostic of $L_{\rm bol}/M$ does not strictly reflect the evolutionary stage of deeply embedded young stellar objects whose envelopes may just be too  compact to impact the $L_{\rm bol}$ on larger scales. An interesting case is G351.58$-$0.4, which has a low $L_{\rm bol}/M$, and hence it is expected to be among the younger sources. However, as it shows mid-infrared emission, it has been classified as an infrared bright source in \citet{Konig2017}. Based on radio observations, \citet{Culverhouse2011} find a spectral index consistent with the criteria for \uchii\ regions from \citet{Wood1989}, suggesting that it hosts already formed high-mass star(s). Nevertheless, consistently with its low $L_{\rm bol}/M$, we find a relatively high $X$(OD), suggesting that the bulk of the gas is chemically young and has not yet been strongly impacted by the star formation process.

Our findings suggest that the OD abundance in the cold dense gas is higher at the onset of the collapse and decreases with time. This is consistent with chemical models predicting that both a higher temperature and stronger radiation field lead to a more efficient destruction of OD. Due to the radiative feedback of newly formed high-mass stars, such conditions  occur towards the more advanced evolutionary stages of high-mass star formation. Furthermore, through reactions with O, CO, and C, the OD molecule is destructed in the dense gas; however, these are rather slow processes \citep{Croswell1985}, suggesting that the lifetime of OD in the cold dense gas is long.  The example of G351.58$-$0.4 suggests that the presence of deeply embedded \uchii\ regions does not have a strong impact on the OD abundance in the outer layers of the clump, instead the overall $L_{\rm bol}/M$ ratio seems more important.

\subsection{OD and HDO: Photodissociation and constraints on the water chemistry}\label{sec:photo}

The formation and destruction of  OH 
(and OD) are tightly linked to the chemical reaction network of water. Since data for the ground-state HDO line is available for roughly half of the sample, for the more evolved sources, we typically compare and correlate here the observed OD and HDO columns in the cold envelope. From the line shapes shown in Fig.\,\ref{fig:od_all}, we already notice a resemblance suggesting that the absorptions originate from similar conditions within the envelope. We used the measured area for the \oht\ OD ($J = 5/2 - 3/2$) line from the line-to-continuum ratio (which is directly proportional to the molecular column density in the case of optically thin emission), and perform the same analysis with the HDO ground-state line. Similarly, as in the case for OD, we fitted two velocity components to the HDO line for W49N and W51e2, and we separated the contribution from the outflow and the  $70$\,\kms\ components, respectively. Fig.\,\ref{fig:od-hdo-corr} shows the comparison between the OD and HDO absorption from the ground-state transitions. Our results suggest that the OD and HDO column density is correlated; however, we have to note that this result is dominated by the source, G351.58$-$0.4, which has by far the largest amount of OD and HDO columns. Unfortunately, due to the lack of HDO observations towards a larger number of sources considered young, we cannot conclude whether such a trend holds. However, a tentative correlation between the HDO and OD column would suggest that OD could be a proxy for HDO, and hence H$_2$O, in the outer envelope. 

A correlation between OD and HDO might be expected since the formation pathways are tightly linked between OH and H$_2$O.  In the cold gas, OH forms together with water as a result of dissociative recombination from H$_3$O$^+$ \citep{Buhr2010,vanDishoeck2014}. At low temperatures, however, the high water abundance in ices favours a significant contribution from grain surface processes as well \citep{Tielens1982,Roberts2002a, Cuppen2010}.
On the other hand, at higher temperatures, OH could form through direct reactions between H$_2$ and O if the energy barrier could be overcome
\citep[e.g.][]{Elitzur1978,Wagner1987,vanderTak2006}. A subsequent reaction with H$_2$ then forms followed by H$_2$O. This reaction would essentially convert all OH to H$_2$O, however, the backward reaction proceeds when a strong UV field is present or a high abundance of atomic hydrogen is prevalent. The destruction of OH and H$_2$O is further differentiated by different reaction channels. While H$_2$O is destroyed by an ion-molecule reaction with C$^+$,  OH is destroyed by a neutral-neutral reaction with N \citep{Wakelam2010}. More importantly, the equilibrium of the fundamental production and destruction channel of OH + H$_2$ {$\rightleftharpoons$} H$_2$O + H depends on the local conditions, such as density and temperature. Altogether, this may lead to differences in the OH and H$_2$O abundances throughout the inner envelope. However, when considering the main production channel for H$_2$O, the same reaction for OD + H$_2$ {$\rightleftharpoons$} HDO + H is likely to be slower \citep[e.g.][]{Thi2010}. Therefore the OD over OH ratio is expected to change once the chemical reaction pathways change in the inner regions. This is expected to lead to a general decrease of HDO/H$_2$O in the hot inner regions, while the OD/OH abundance is expected to increase, and these differences should become more pronounced with time \citep{Coutens2014}.

Based on our radiative transfer modelling of the OH and OD lines, we constrained the deuterium fractionation for OH in the outer envelope. The best estimates are obtained for G34.26+0.15 and G351.58$-$0.4, which are 4.8$\times10^{-3}$ and  {2.5}$\times10^{-2}$, respectively. We obtain an upper limit of the order of 1.5$\times$10$^{-2}$ for G35.20$-$0.7 and a lower limit of 2$\times$10$^{-2}$ for G327.29$-$0.6, which are all sources with low $L_{\rm bol}/M$.
Although we do not have a large sample of sources with deuterium fractionation for both OH and H$_2$O, the values collected in Table\,\ref{tab:results} show that the OD/OH and HDO/H$_2$O values are similar within an order of magnitude. The source G351.58$-$0.4 shows, however, somewhat larger deuterium fractionation for OD compared to HDO. 

While the inner regions are expected to be greatly impacted by the photodissociation of H$_2$O, we cannot put strong constrains on the deuterium fractionation of OD in the hot inner regions here. Our observations exclude, however, a stronger fractionation at regions of a few thousands of au for G34.26+0.15. Observations of higher J transitions of OD could put more constrains on the OD/OH fractionation in the inner region, although large optical depths may prohibit these observations.

\section{Conclusions}
We report the detection of the $^2\Pi_{3/2}$ OD ($J = 5/2 - 3/2$) line at 1.3\,THz with the SOFIA/GREAT instrument towards a sample of Galactic high-mass star forming regions. We estimate the OD column densities in a range of $N$(OD)$=4-70\times10^{12}$\,cm$^{-2}$ for these sources, where the highest OD column densities are seen towards the youngest clumps of the sample. 

We performed a detailed radiative transfer modelling for one of the best known source of the sample, G34.26+0.15. Our results show that the $^2\Pi_{3/2}$ OD ($J = 5/2 - 3/2$) absorption originates from the dense and cold outer layers of the envelope. We put constraints on the deuterium fractionation within the inner regions, where, due to sublimation and higher temperatures, gas-phase chemical formation pathways are expected to dominate the chemistry. Although the  limits obtained from these models are not stringent,   they allow us to exclude a strong enhancement of deuterium fractionation towards the inner envelope. 

We estimate the OD molecular abundance in the outer envelope for nine sources of the sample. A correlation with the $L_{\rm bol}/M$ ratio - which is used as a proxy for evolutionary stage - shows a tentative slow decrease in the OD abundance with time. This is consistent with chemical models of the cool, dense gas phase, where the main formation of OD proceeds through rapid exchange reactions leading to a strong enhancement of deuterium fractionation.  The OD molecule is then destroyed by slow reactions taking place in the outer envelope. Our results suggest that radiation from the deeply embedded heating sources at this stage does not strongly impact the conditions in the envelope.

We find a deuterium fractionation for OD of  $\sim$0.5\% towards G34.26+0.15 in the outer envelope, which is consistent with estimates from the HDO/H$_2$O ratio. We find a weak correlation between the OD and HDO column in a sample of 6 sources for which  ancillary data were available. This suggests that OD could be a proxy for HDO and thus H$_2$O in these sources.

\bibliographystyle{aa}
\bibliography{od_sofia}

\begin{acknowledgements}
We thank the referee for the careful reading of the manuscript and constructive suggestions that improved the paper.
Based on observations made with the NASA/DLR Stratospheric Observatory for Infrared Astronomy. SOFIA Science Mission Operations are conducted jointly by the Universities Space Research Association, Inc., under NASA contract NAS2-97001, and the Deutsches SOFIA Institut under DLR contract 50 OK 0901. T.Cs. acknowledges financial support from the French State in the framework of the IdEx Université de Bordeaux Investments for the future Program.
The development and operation of GREAT is financed by resources from the participating institutes,  by the Deutsche Forschungsgemeinschaft (DFG) within the grant
for the Collaborative Research Center 956 as well as by the
Federal Ministry of Economics and Energy (BMWI) via the
German Space Agency (DLR) under Grants 50 OK 1102,
50 OK 1103 and 50 OK 1104.
\end{acknowledgements}

\appendix
\section{Model results for G351.58$-$0.4}\label{app:g351}

 {In Fig.\,\ref{fig:g351}, we show the results of our RATRAN modelling for G351.58$-$0.4. As for the other sources, these models assume a constant abundance profile for all species. While the fit for the OD $^2\Pi_{3/2}$ J=5/2-3/2 transition at 1390.61\,GHz is satisfactory, the OH $^2\Pi_{1/2}$ J=3/2-1/2 transition is not well reproduced by this model. As suggested by the NH$_3$ ($J=3_{2+}-2_{2-}$) line profile from \citet{Wyrowski2016}, the blueshifted component of a broader absorption component is apparent, which also seems to impact the excited-state OH line profiles. Furthermore, our assumption on the constant OH abundance profile may also be inadequate for this source.}

    \begin{figure*}[!ht]
   \centering
      \includegraphics[width=\linewidth]{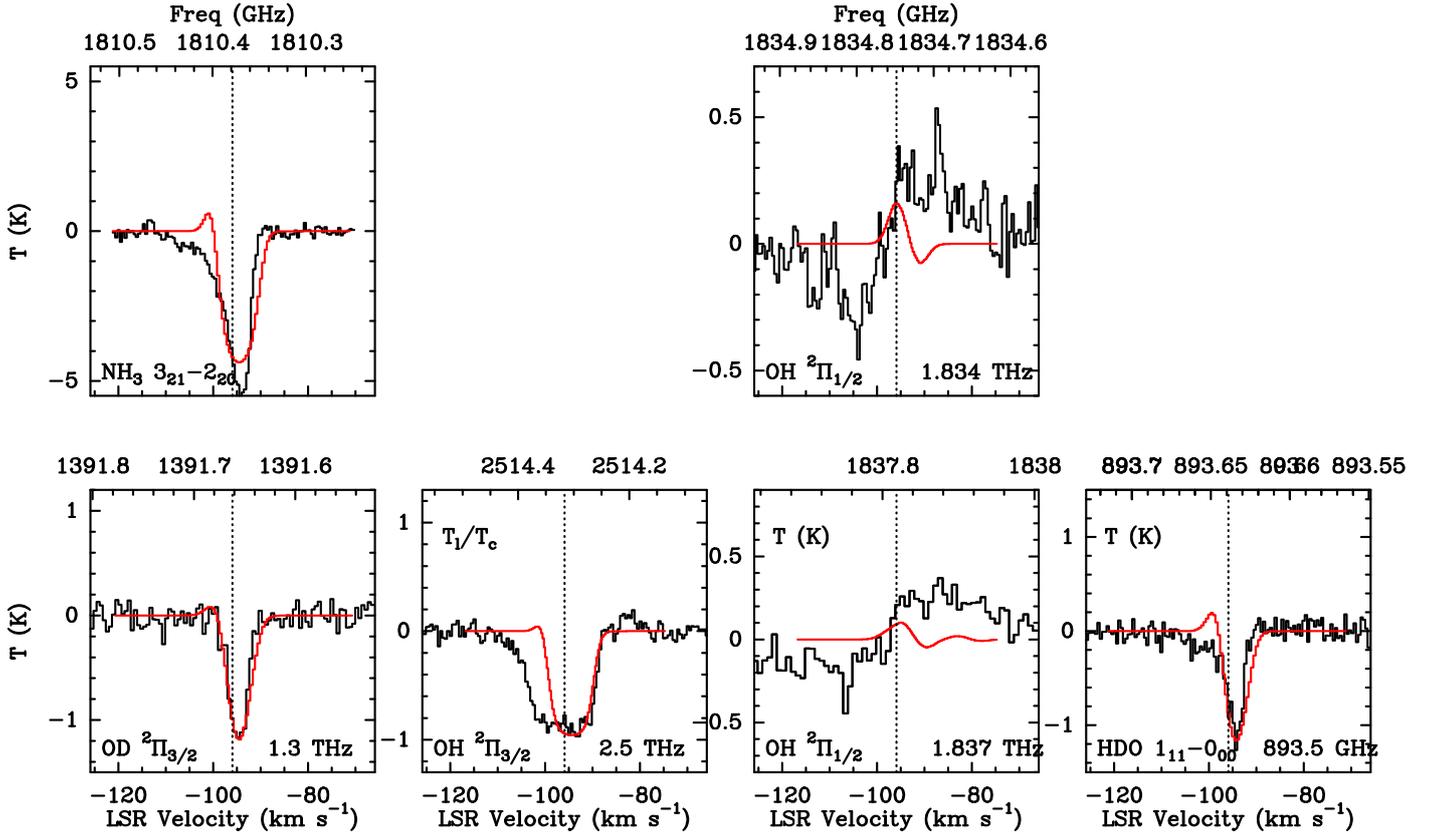}
      \caption{
       {Same as Fig.\,\ref{fig:g34}, but for G351.58$-$0.4.}
      The observed spectra are shown in black, red line shows the fitted RATRAN model. Dotted line marks the $v_{\rm lsr}$ of the source.  {The panel for the  $^2\Pi_{3/2}$ J=5/2-3/2 transition at 2514\,GHz shows the line-to-continuum ratio ($T_{\rm l}/T_{\rm c}$), all other panels show the line temperature in Kelvins.}
      }
              \label{fig:g351}%
    \end{figure*}
   \section{Comparison with the NH$_3$ ($J=3_{2+}-2_{2-}$) 1.8 THz absorption}\label{app:nh3}
   
    {We compare the line profiles of the OD $^2\Pi_{3/2}$ J=5/2-3/2 line with the NH$_3$ ($J=3_{2+}-2_{2-}$) line in Fig.\,\ref{fig:od-nh3}. For better visibility, the NH$_3$ lines are scaled. We notice that the line profiles are remarkably similar, showing very similar rest velocities. Towards the sources G327.29$-$0.6 and G351.58$-$0.4, absorption from a blueshifted broad velocity component, likely due to molecular outflows, is visible in the NH$_3$ line, while it is absent in the ground-state OD line.
   }

   \begin{figure*}[!h]
   \centering
   \includegraphics[width=5.9cm]{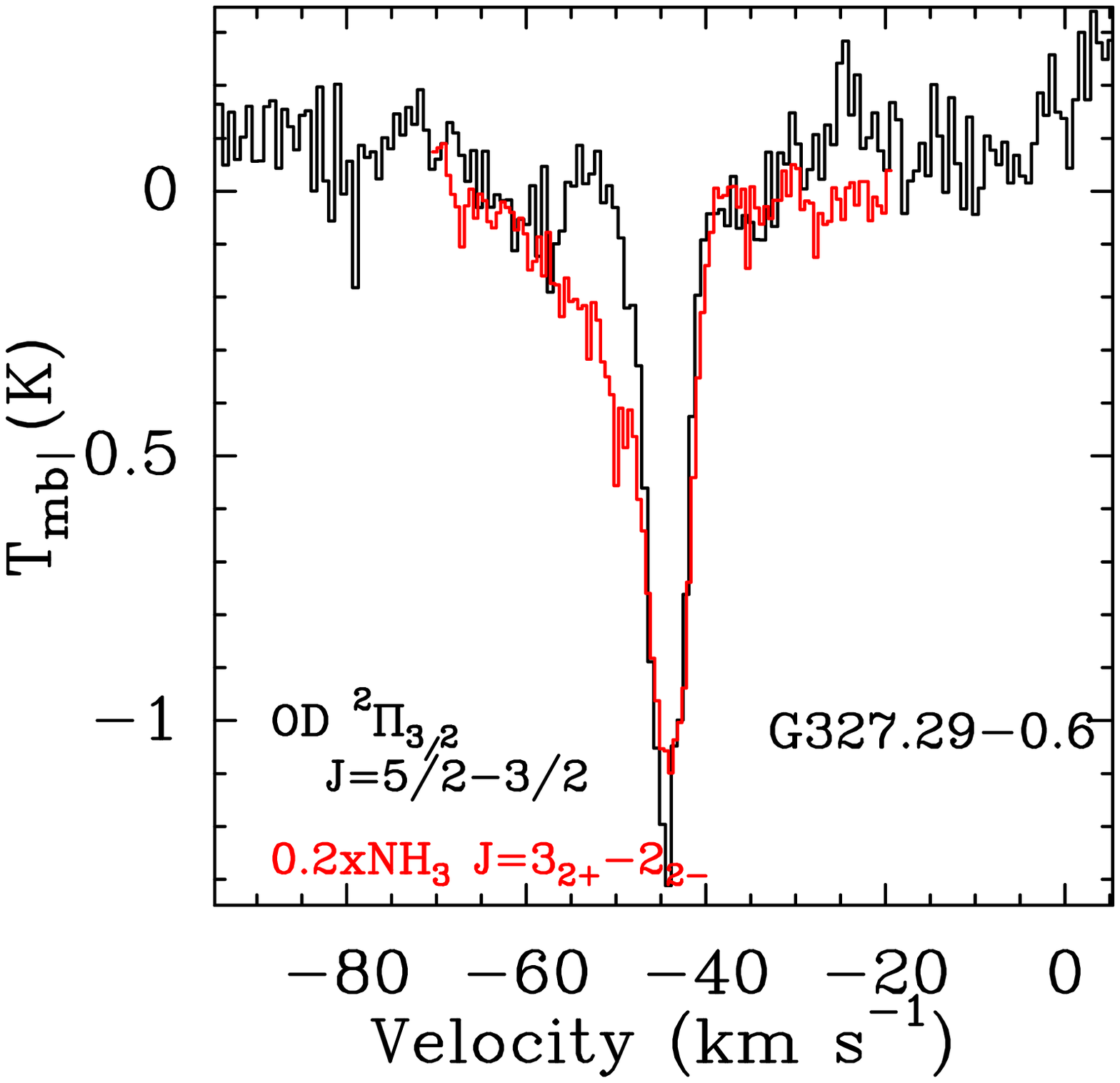}
   \includegraphics[width=5.9cm]{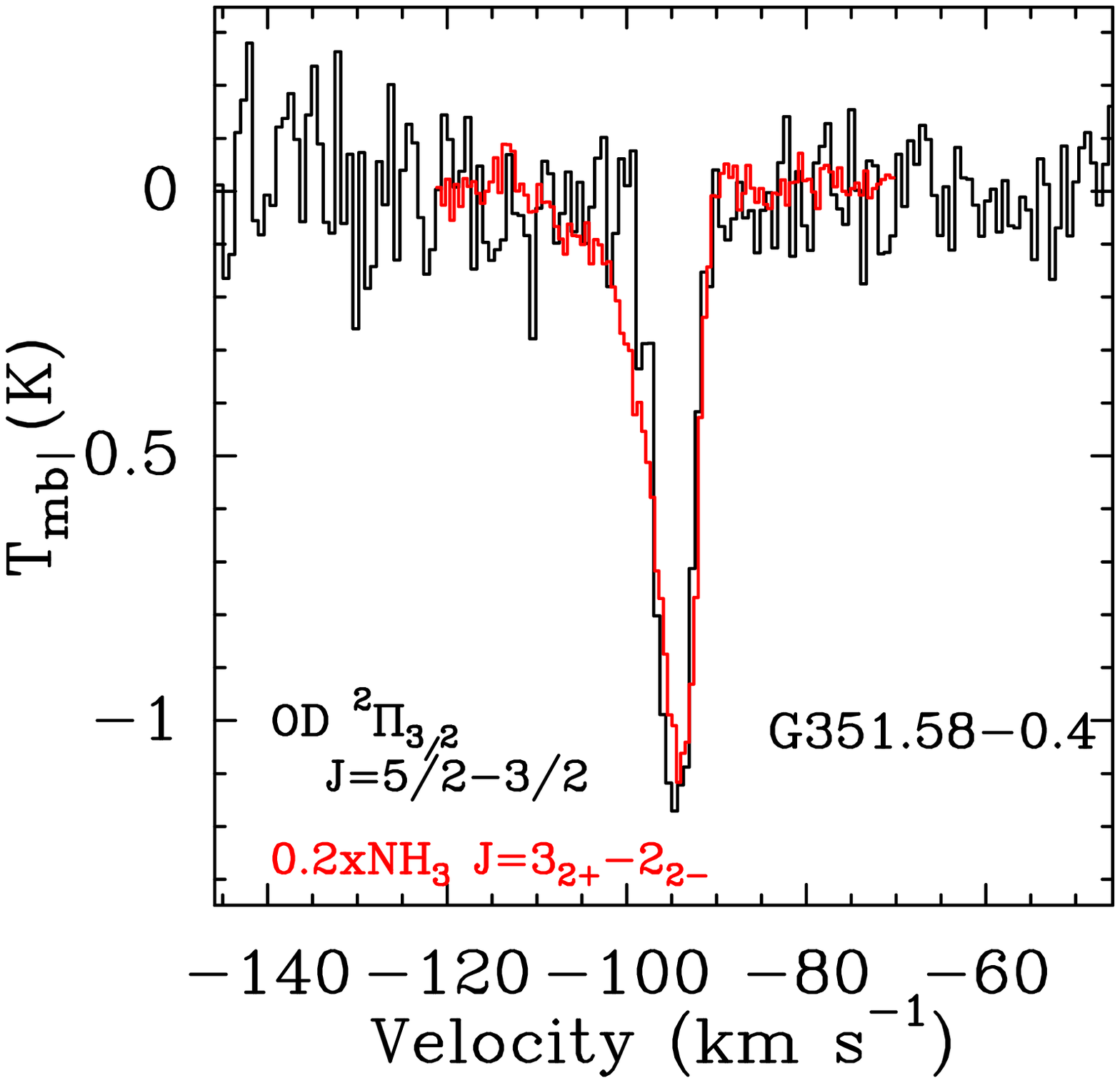}
   \includegraphics[width=5.9cm]{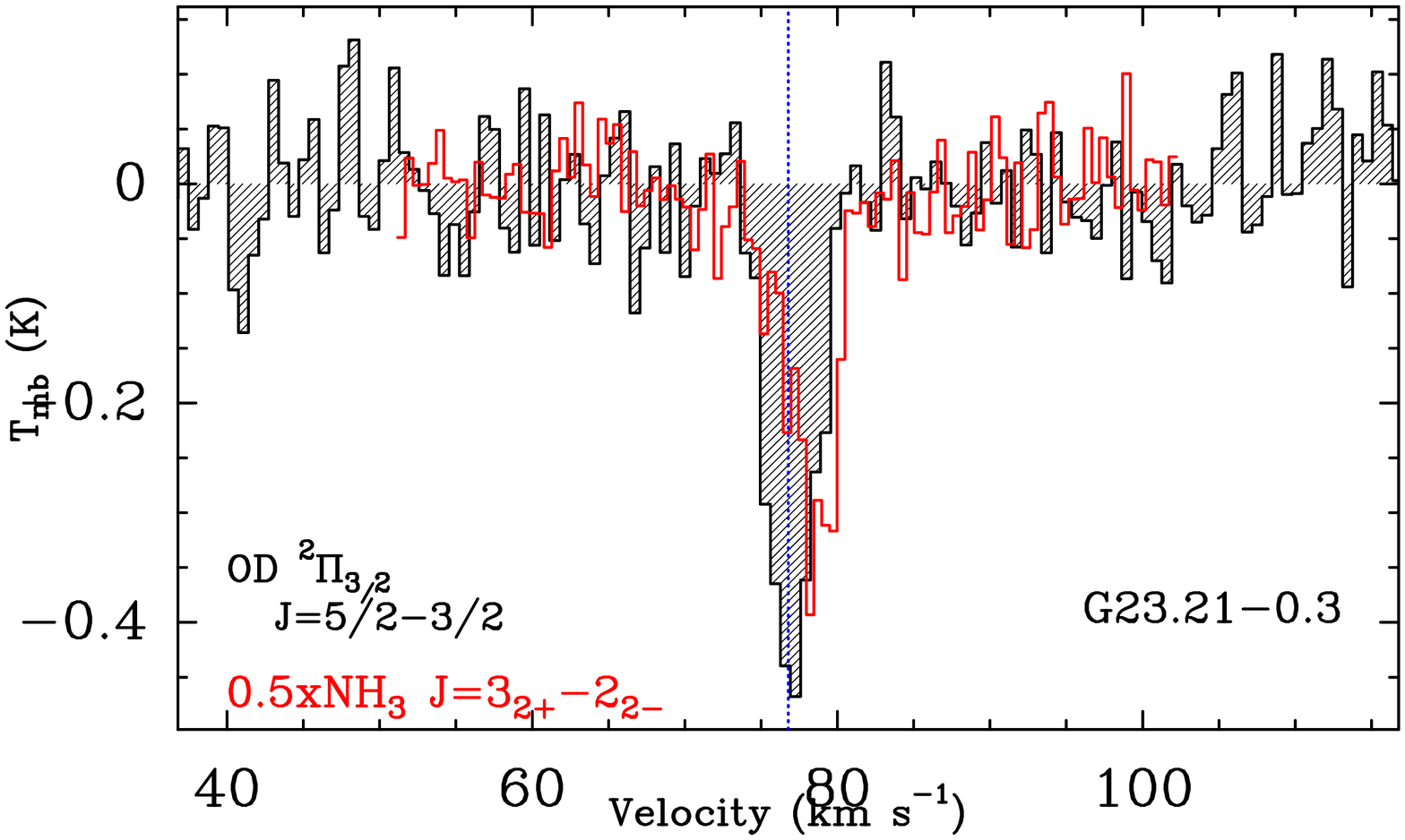}
   \includegraphics[width=5.9cm]{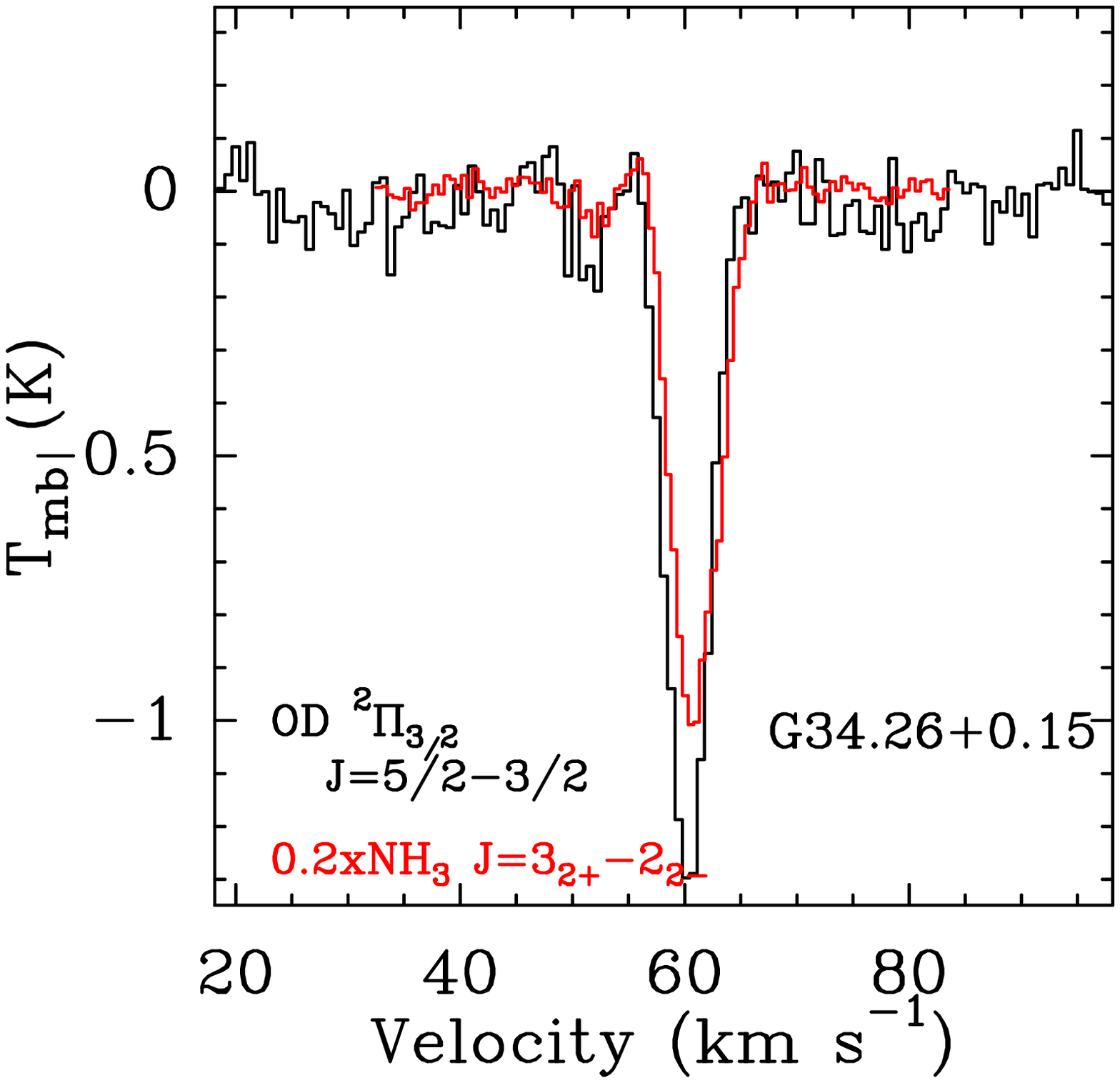}
   \includegraphics[width=5.9cm]{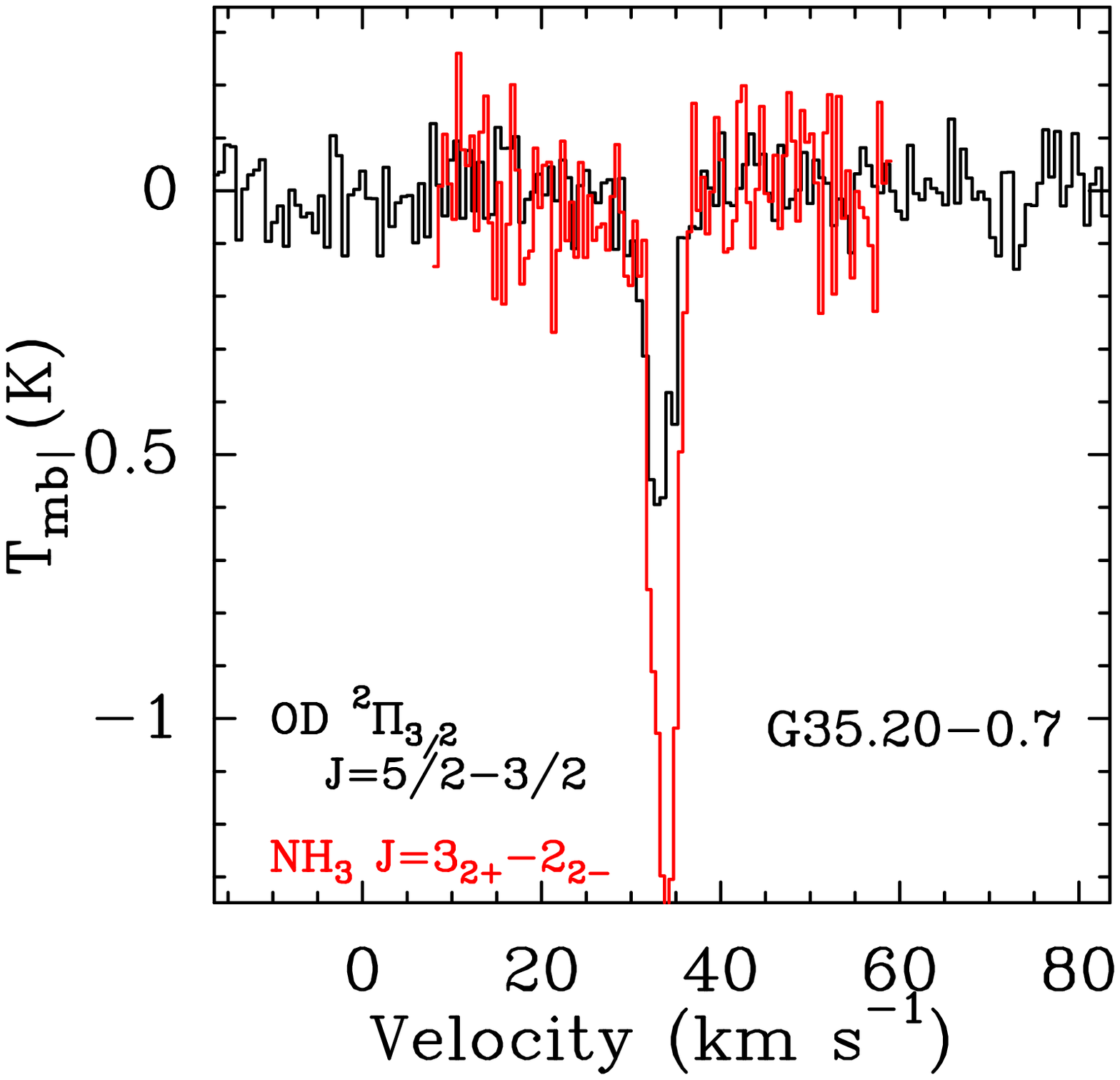}
   \includegraphics[width=5.9cm]{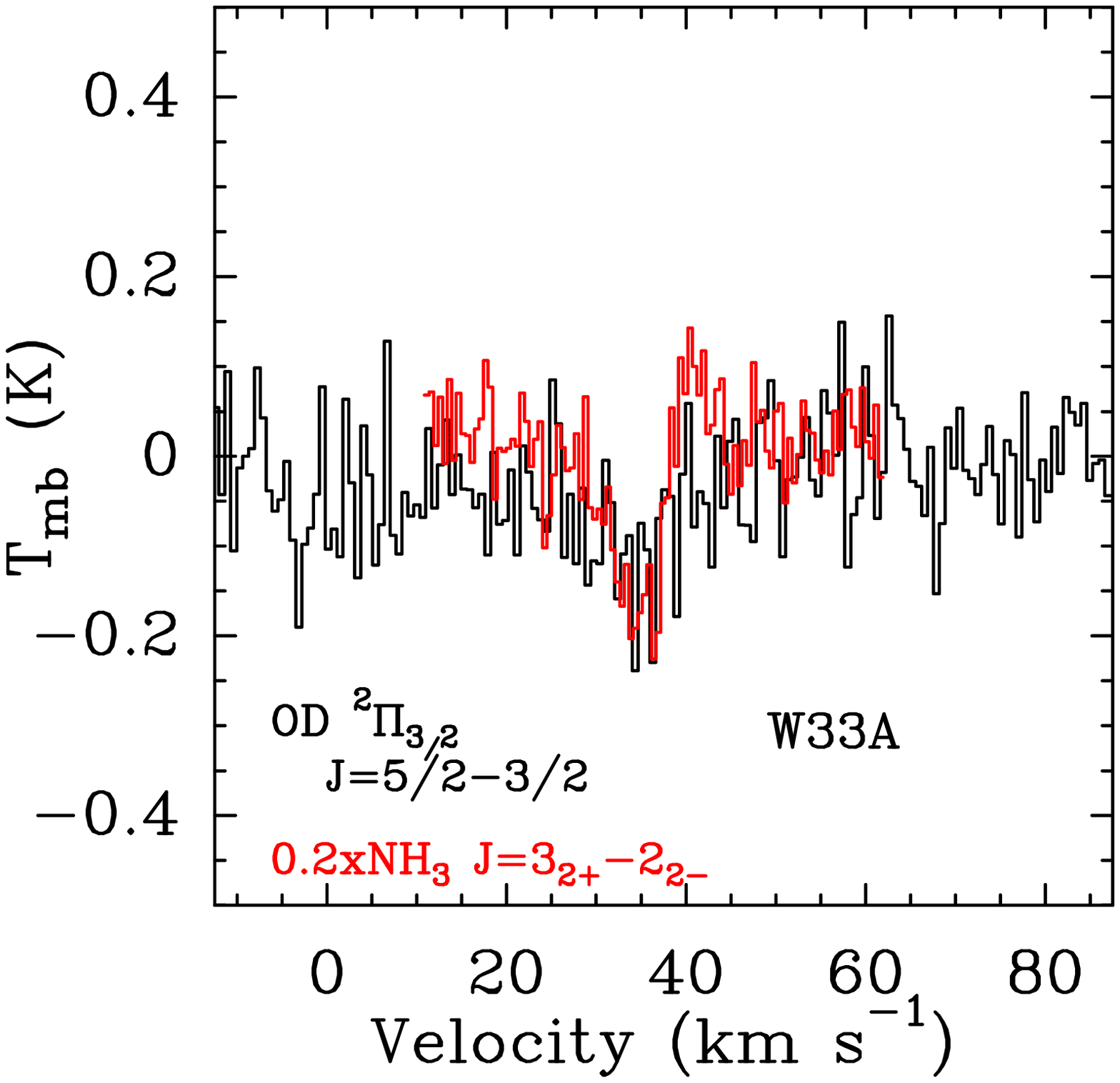}
   \includegraphics[width=5.9cm]{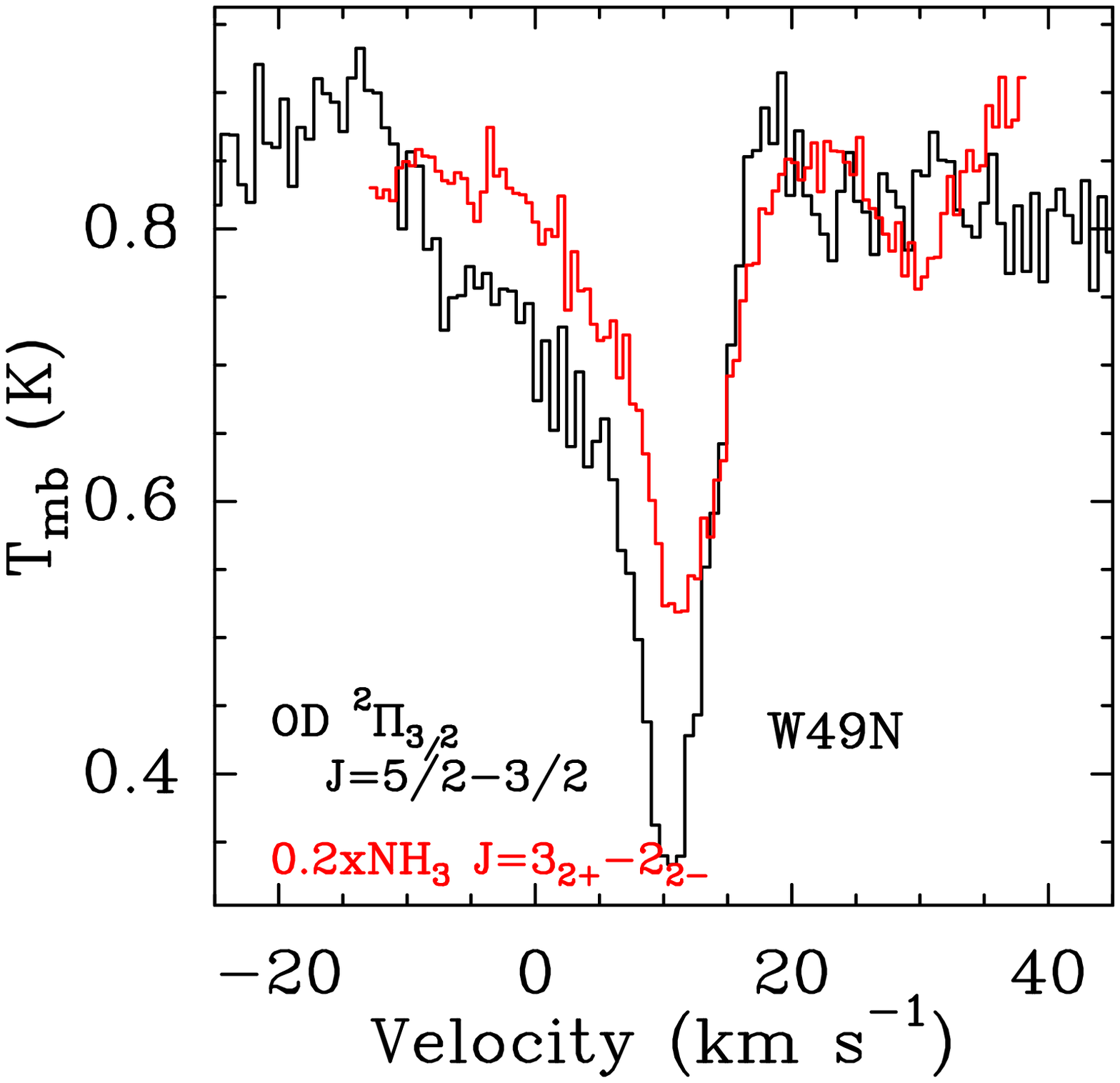}

      \caption{Comparison between the OD $^2\Pi_{3/2}$ ($J=5/2-3/2$) transition (\textit{black}) and the NH$_3$ ($J=3_{2+}-2_{2-}$) transition (\textit{red}).
      }
              \label{fig:od-nh3}%
    \end{figure*}
 
\section{Removal of instrumental artefacts from the
$^{18}\mathrm{OH}\,^2\Pi_{3/2}$ ground-state line}
\label{app:18oh}
In supra-THz heterodyne spectroscopy, local oscillator (LO) power is precious,
and transmission losses in the injection of the generated signal are to be kept
at the minimum. The $^{18}\mathrm{OH}\,^2\Pi_{3/2}$ ground-state line was obtained in 2013, when the LO
signals of GREAT \citep{Heyminck2012} were injected by means of Martin-Pupplet
diplexers \citep[e.g.][]{Lesurf1981}. Depending on the tuning, standing waves may arise, causing sinusoidal baselines that are further modulated by the limited frequency passband of the diplexer. Moreover, beyond standing wave features, gain fluctuations
can originate in the mixing or subsequent amplification stages. If they are faster
than the occurrence of calibration scans, they further deteriorate the baseline
quality, especially in observations towards far-infrared, strong continuum sources. These effects together impact
the tuning of the Ma band to the
$^{18}\mathrm{OH}\,^2\Pi_{3/2}$ line.

By their nature, these standing waves cannot be removed by a simple Fourier filter,
unlike their equivalents occasionally arising between the mixer and the telescope's
secondary mirror. The fitting of high-order polynomial baselines ($\sim$10) was found to introduce spurious line profiles at the systemtic velocity and is, likewise, to be avoided. Adjusting sinusoidal waveforms, although numerically somewhat more stable, lacks
objectiveness as well. In contrast, differential total power spectra, extracted from the off-source phases of subsequent chop-and-nod sub-scans and typically separated by $\sim 1$\,min, were observed to exhibit similar baseline distortions while being void of the astronomical signal \citep[][further references therein]{Higgins2020}. In the stable part of the spectral bandpass the aforementioned artefacts are linear in
nature, which suggests the necessity to use a linear combination of such differential off-source
total power spectra (hereafter $\mathbf{a}_\mathrm{j},\,j=1,2,\cdots,n$) to fit the
residual baseline apparent in the averaged on-off spectrum, denoted $\mathbf{b}$. All spectra are represented in an $m$ dimensional vector space, where $m$ is the number of spectral channels considered. For this purpose, we minimised the quantity:
\begin{equation}
\chi^2 = | A \cdot \mathbf{p} - \mathbf{b}|^2,
\label{eq:chi2}
\end{equation}
where the columns of the $m\times n$ matrix A are the spectra $\mathbf{a}_\mathrm{j}$, and where the coefficients of the linear combination minimizing $\chi^2$ are denoted as a vector $\mathbf{p} = (p_1,p_2,\cdots,p_n)^\mathrm{T}$. Since $m>n$, the problem to be solved is over-determined; yet, a direct minimisation using normal equations is often numerically unstable, because ambiguous combinations of coefficients $\mathbf{p}$ can occur, causing the Jacobian to become singular. However,
the solution vector $\mathbf{p}$ can be obtained from the easily invertible singular
value decomposition (SVD) of matrix A, providing a more stable method to minimise $\chi^2$ \citep[e.g. \textit{\emph{Numerical Recipes}},][]{Press1992}.
Our baseline fit can then be constructed as $\mathbf{b} = \sum_{j=1}^{n} p_\mathrm{j} \mathbf{a}_\mathrm{j}$.
\\[1.5ex]
In practice, we do not directly use the spectra $\mathbf{a}_\mathrm{j,}$ we used
smoothed fitting functions instead; this is to avoid excess noise otherwise introduced
by the baseline removal. Third order wavelet transforms proved to
provide baseline fits with the smallest $\chi^2$. Other similarly smoothed
variants may yield satisfactory results alike, while higher orders bear the risk of
fitting the spectral line as a baseline feature even though the spectra
$\mathbf{a}_\mathrm{j}$ do not contain the astronomical signal. As a safeguard against this,
we use a weighted fit, modifying Eq.~\ref{eq:chi2}:
\begin{equation}
\chi^2 = \left| \left(\mathrm{diag}(w_\mathrm{j})\right) \cdot \left (A \cdot \mathbf{p} - \mathbf{b} \right)\right|^2,\, j=1,\cdots,m,
\label{eq:wchi2}
\end{equation}
with weights
\begin{equation}
w_\mathrm{j} = 1-\exp\left(-4 \ln{2}\cdot \left(\frac{(\upsilon_\mathrm{j}-\upsilon_0)}{\Delta_\upsilon}\right)^2\right),
\label{eq:weight}
\end{equation}
centred at $\upsilon_0 = 60\,\mathrm{km\,s^{-1}}$. The exact choice of the width $\Delta_\upsilon$ is not critical as long as the putative $^{18}\mathrm{OH}$
absorption line is covered, yet leaving enough spectral baseline on either side (i.e.,
$\Delta_\upsilon = 10$ to $15\,\mathrm{km\,s^{-1}}$). For the computations we use
\textit{Numerical Recipes} routines \textit{svdcmp} and \textit{svbksb} to perform the
SVD and to calculate $\mathbf{p}$, respectively.
\begin{figure}[!h]
\resizebox{\columnwidth}{!}{\includegraphics{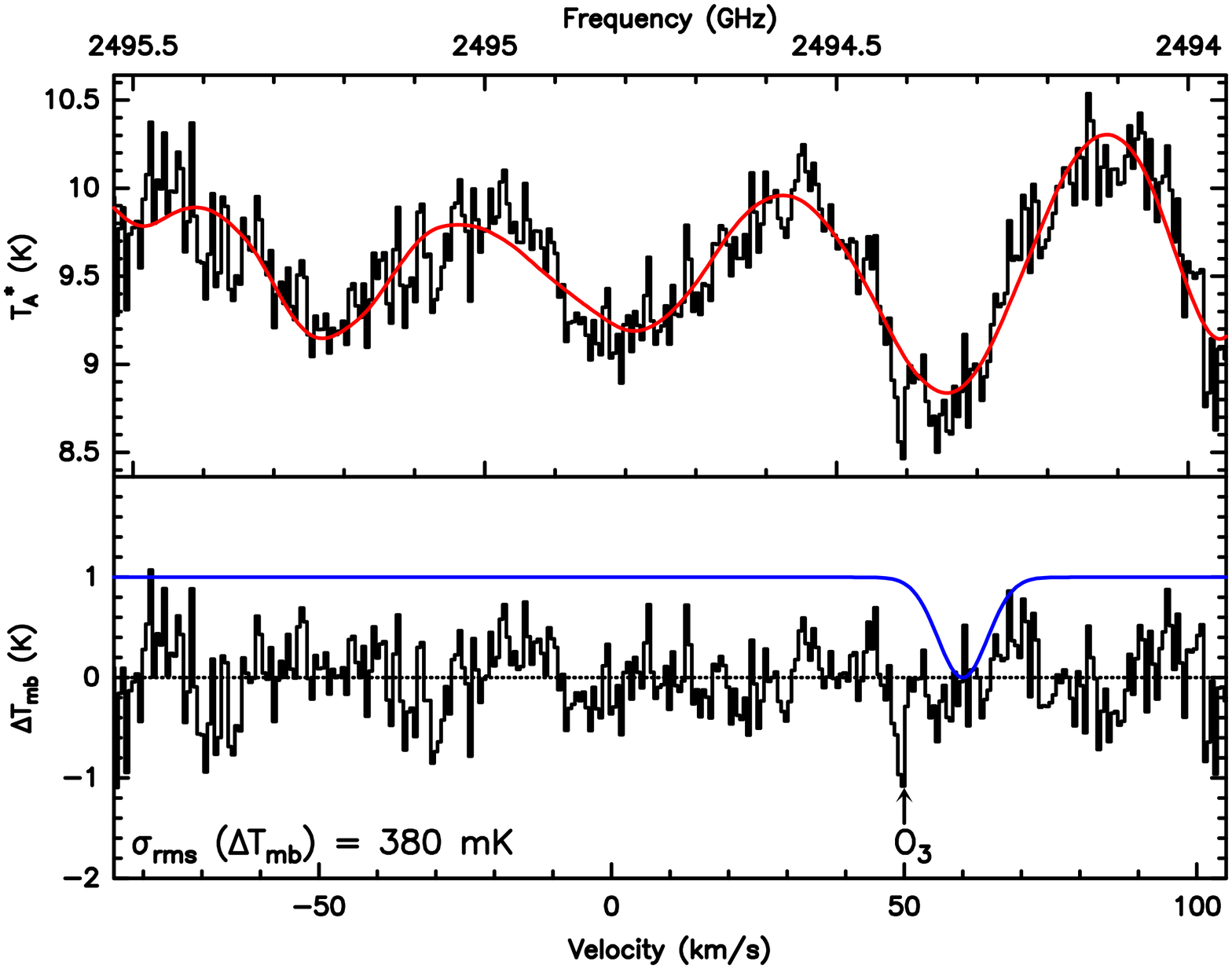}}
\caption{Spectrum of $^{18}\mathrm{OH}\,^2\Pi_{3/2}$ before (top)
and after (bottom) removal of the SVD derived baseline (red line, top) and
transmission correction. The relatively narrow kernel of a telluric $\mathrm{O}_3$ line is deliberately left uncorrected. The blue line represents the weight profile used in the $\chi^2$ minimisation.
\label{fig:c1}
}
\end{figure}

The application of the SVD technique to the baseline removal from the
$^{18}\mathrm{OH}\,^2\Pi_{3/2}$ line is demonstrated in Fig.~\ref{fig:c1}.
After iterative Hanning smoothing to $0.6\,\mathrm{km\,s^{-1}}$ channel separation, the resulting spectrum features a sensitivity of $\sigma_\mathrm{rms}(T_\mathrm{A}^*) = 207$~mK prior to transmission correction, which equals, within errors, the radiometric noise pertaining to the measured single-sideband system temperature in the relevant part of the bandpass, 6070~K. The obtained spectrum therefore qualifies to define a lower limit to the $^{16}\mathrm{OH}/^{18}\mathrm{OH}$ ratio (Sect.\,\ref{sec:ratran} and Fig.\,\ref{fig:g34}).

\section{Comparison of RATRAN models with different collisional rate coefficients}
\label{app:od_coeff}

In Sect.\,\ref{sec:ratran}, we used RATRAN to model the OD absorption, and adopted the collisional rate coefficients of  OH  \citep{Klos2017} also for OD. However, recently, collisional rate coefficients for OD with H$_2$ became available \citep{Dagdigian2021} that also include the hfs splitting. To test the validity of our initial approach, we compared our best fit model for the OD 1391\,GHz line (Sect.\,\ref{sec:ratran_g34}) for G34.26$+$0.15 both using the OH coefficients from  \citet{Klos2017} adapted for OD, and the collisional rate coefficients of OD including the hfs structure from  \citet{Dagdigian2021}. We show this comparison in Fig.\,\ref{fig:coll_comp} and see that the difference between the models (red lines) is marginal.

\begin{figure}[!h]
\resizebox{\columnwidth}{!}{\includegraphics{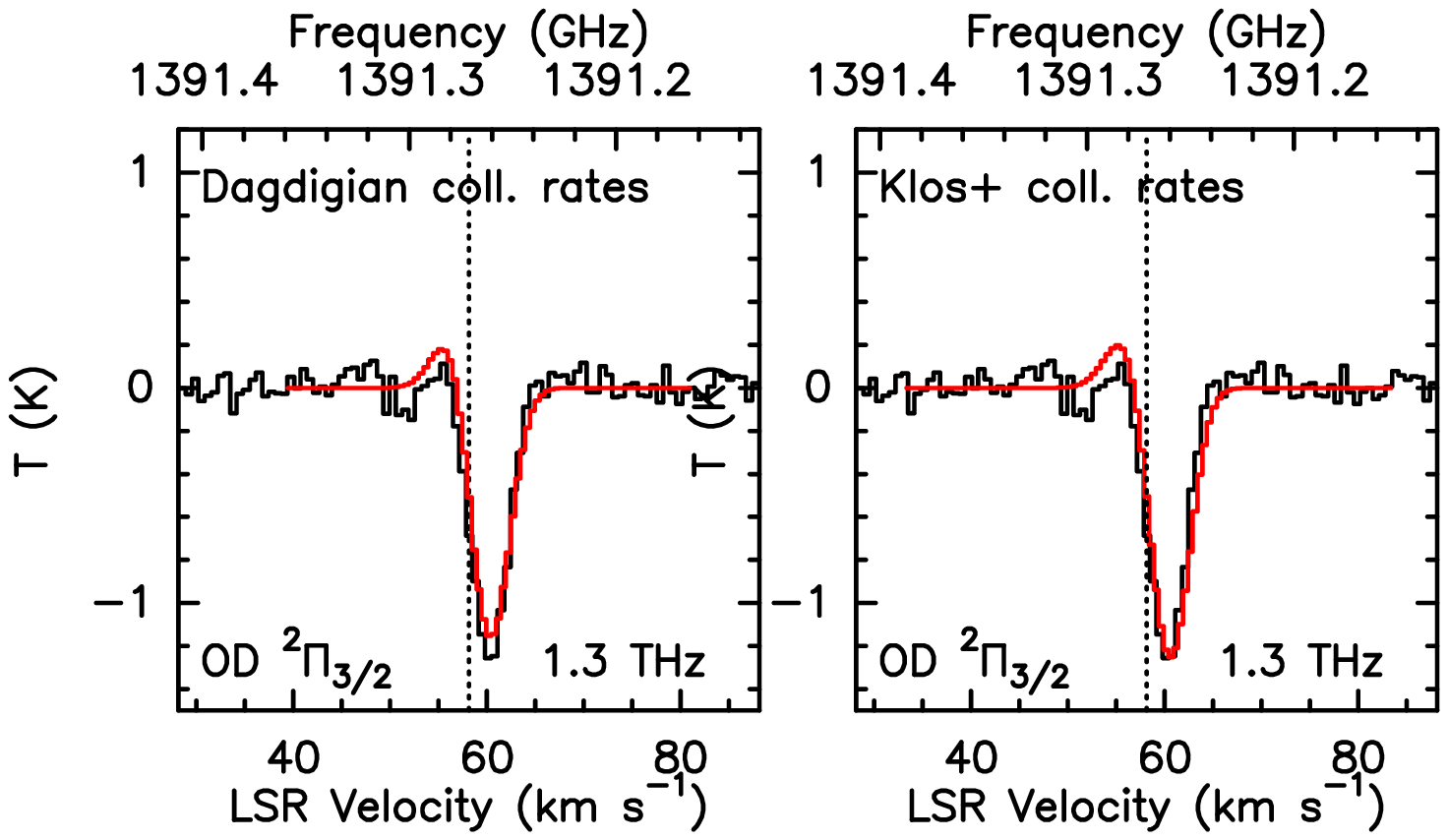}}
\caption{Comparison of our RATRAN models for the OD 1391\,GHz line using the collisional rate coefficients for OD from \citet{Dagdigian2021} (left) and the collisional rate coefficients from OH adapted for OD from \citet{Klos2017} (right).}
\label{fig:coll_comp}
\end{figure}

\end{document}